\pdfoutput=1 %for arxiv
\documentclass[a4paper,twocolumn,fleqn,usenatbib]{mnras}
\usepackage[T1]{fontenc}
\usepackage{ae,aecompl}
\usepackage{caption}

\usepackage{graphicx}	% Including figure files
\usepackage{amsmath}	% Advanced maths commands
\usepackage{amssymb}	% Extra maths symbols

\newcommand{\de}{\mathrm{d}}

\newcommand{\ramses}{\texttt{RAMSES}}
\newcommand{\phew}{\texttt{PHEW}}

\newcommand{\exc}{\texttt{exclusive}}
\newcommand{\inc}{\texttt{inclusive}}
\newcommand{\sad}{\texttt{strictly bound}}
\newcommand{\nosad}{\texttt{loosely bound}}

\newcommand{\msol}{M_{\sun}}

\title[ACACIA]{{ACACIA}: a new method to produce on-the-fly merger trees in the RAMSES code }

\author[Mladen Ivkovic]{
    Mladen Ivkovic,$^{1}$$^{2}$\thanks{E-mail: mladen.ivkovic@epfl.ch}
    Romain Teyssier$^{3}$
    \\
    $^{1}$Laboratoire d'Astrophysique, \'Ecole Polytechnique F\'ed\'erale de Lausanne, 1290 Versoix, Switzerland\\
    $^{2}$Observatoire de Gen\`eve, Universit\'e de Gen\`ve, Chemin Pegasi 51, 1290 Versoix, Switzerland\\
    $^{3}$Institute for Computational Science, University of Zurich, 8057 Zurich, Switzerland\\
}

\date{Accepted XXX. Received YYY; in original form ZZZ}

\pubyear{2021}

\begin{document}
    \label{firstpage}
    \pagerange{\pageref{firstpage}--\pageref{lastpage}}
    \maketitle
    
    \begin{abstract}
  The implementation of \texttt{ACACIA}, a new algorithm to generate
  dark matter halo merger trees with the Adaptive Mesh Refinement
  (AMR) code \ramses, is presented.  The algorithm is fully parallel
  and based on the Message Passing Interface (MPI). As opposed to most
  available merger tree tools, it works on the fly during the course
  of the N body simulation.  It can track dark matter substructures
  individually using the index of the most bound particle in the
  clump.  Once a halo (or a sub-halo) merges into another one, the
  algorithm still tracks it through the last identified most bound
  particle in the clump, allowing to check at later snapshots whether
  the merging event was definitive, or whether it was only temporary,
  with the clump only traversing another one.  The same technique can
  be used to track orphan galaxies that are not assigned to a parent
  clump anymore because the clump dissolved due to numerical
  over-merging.  We study in detail the impact of various parameters
  on the resulting halo catalogues and corresponding merger histories.
  We then compare the performance of our method using standard
  validation diagnostics, demonstrating that we reach a quality
  similar to the best available and commonly used merger tree tools.
  As a proof of concept, we use our merger tree algorithm together
  with a parametrised stellar-mass-to-halo-mass relation and generate
  a mock galaxy catalogue that shows good agreement with observational
  data.
\end{abstract}


%%%%%%%%%%%%%%%%%%%%%%%%%%%%%%%%%%%%%%%%%%%%%%%%%%

%%%%%%%%%%%%%%%%% BODY OF PAPER %%%%%%%%%%%%%%%%%%

    %==================================
\section{Introduction}
%==================================

Mock galaxy catalogues generated using N-body or hydrodynamical
simulations are important tools for extragalactic astronomy and
cosmology.  They are used to test current theories of galaxy
formation, to explore systematic and statistical errors in large scale
galaxy surveys and to prepare analysis codes for future dark energy
mission such as Euclid or LSST.  There is a large variety of methods
to generate such mock galaxy catalogues.  The most ambitious line of
products is based on full hydrodynamical simulations, where dark
matter, gas, and star formation are directly simulated
\citep[e.g.][]{duboisDancingDarkGalactic2014,
  khandaiMassiveBlackIISimulationEvolution2015,
  vogelsbergerPropertiesGalaxiesReproduced2014,
  schayeEAGLEProjectSimulating2015}.
The intermediate approach is based on semi-analytic modelling
(hereafter SAM) \citep[e.g.][]{SA-white,
  SA-durham, SA-Somerville, SA-Kaufmann,
  kangSemianalyticalModelGalaxy2005,crotonManyLivesActive2006} for
which galaxy formation physics, although simplified, is still at the
origin of the mock galaxy properties. Finally, the simplest and most
flexible approach is based on a purely empirical modelling of galaxy
properties, sometimes called Halo Occupation Density (HOD hereafter)
\citep[e.g.][]{HOD-Seljak, HOD-Berlind,
  peacockHaloOccupationNumbers2000,
  bensonNatureGalaxyBias2000,wechslerGalaxyFormationConstraints2001,
  scoccimarroHowManyGalaxies2001}.
The last two techniques (SAM and HOD) both require the complete
formation history of dark matter
haloes, and possibly their sub-haloes. This formation history is
described by halo `\emph{merger trees}'
\citep{roukemaSpectralEvolutionMerging1993,
  roukemaFailureSimpleMerging1993, laceyMergerRatesHierarchical1993}.
Accurate merger trees are essential to obtain realistic mock galaxy
catalogues, and constitute the backbone of SAM and HOD models.

The advantage of using SAM and HOD techniques to generate mock galaxy
catalogues is that one does not need to model explicitly the gas
component, but only the dark matter component.  The corresponding
N-body simulations are commonly referred to as `\emph{dark matter
only}' (DMO) simulations.  With growing processing power, improved
algorithms and the use of parallel computing tools and architectures,
larger and better resolved DMO simulations are becoming possible.  The
current state-of-the-art is the Flagship simulation performed for the
preparation of the Euclid mission \citep{PKDGRAV} and featured 2
trillion dark matter particles.  Such extreme simulations make
post-processing analysis tool such as merger tree algorithms
increasingly difficult to develop and to use, mostly because of the
sheer size of the data to store on disk and to load up later from the
same disk back into the processing unit memory.  In some extreme
cases, the amount of data that needs to be stored to perform a merger
tree analysis in post-processing is simply too large.  Storing just
particle positions and velocities in single precision for trillions of
particles requires dozens of terabytes per snapshot.  Another issue is
that most modern astrophysical simulations are executed on large
supercomputers which offer large distributed memory.  Post-processing
the data they produce may also require just as much memory, so that
the analysis will also have to be executed on the distributed memory
infrastructures as well.  The reading and writing of such vast amount
of data to a permanent storage remains a considerable bottleneck,
particularly so if the data needs to be read and written multiple
times.  One way to reduce the computational cost is to include
analysis tools like halo-finding and the generation of merger trees in
the simulations and run them ``\textit{on the fly}'', i.e. run them
during the simulation, while the necessary data is already in memory.

The main motivation for this work is precisely the necessity for such
a merger tree tool for future ``beyond trillion particle''
simulations. While many state-of-the-art N-body simulation codes include
structure finders that are run on-the-fly, codes like \texttt{Gadget4}
\cite{springelSimulatingCosmicStructure2021a} who are able to build
merger trees on-the-fly are still a relative rarity. It is crucial that
multiple, distinct, codes have the capacity to do this to provide the
possibility to cross-check results and their convergence.
To this end, a new algorithm that we named
\texttt{ACACIA} was designed to work on the fly within the parallel
AMR code \ramses.  One novel aspect of this work is the use of the
halo finder \phew\ \citep{PHEW} for the parent halo catalogue.
Different halo finders have been shown to have a strong impact on the
quality of the resulting merger trees \citep{SUSSING_HALOFINDER}.
\phew\ falls into the category of ``watershed'' algorithms that are
not so common in the cosmological halo finding literature.  This type
of algorithm assigns particles (or grid cells) to density peaks above
a prescribed density threshold and according to the so-called
``watershed segmentation'' of the negative density field.

This paper is structured as follows.  In Section \ref{chap:phew}, a
brief description of the \phew\ halo finder and its new particle
unbinding method is given. Section \ref{chap:making_trees} describes
some common difficulties that arise when making merger trees and the
way we address them in our merger algorithm \texttt{ACACIA}, which is
ultimately described in detail in Section \ref{chap:my_code}.
Section \ref{chap:tests} shows test results to
determine what parameters give the best results.  Using the halo
catalogue and its corresponding merger tree generated on the fly by a
cosmological N-body simulation, we use the stellar-mass-to-halo-mass
(SMHM) relation from \cite{Behroozi} to produce a mock galaxy
catalogue. We analyse in Section~\ref{chap:mock_catalogues} the
properties of our mock galaxy catalogue and show that the introduction
of orphan galaxies improve the comparison to observations considerably.
Finally, a detailed comparison with the other halo finding and
tree-building algorithms presented in \citet{SUSSING_HALOFINDER} is
given in Appendix \ref{app:performance_comparison}.

    %=====================================================================
\section{Halo Finding and Particle Unbinding}\label{chap:phew}
%=====================================================================

Halo finding plays a central role in the exploitation of N-body
simulations. Paradoxically, a unique definition of what is a halo or a
sub-halo has never been adopted so far.  The current state of affairs
in the halo finding business is quite the opposite, with a multitude
of definitions emerging over the last decades, each definition
corresponding to a different halo finding algorithm.  The Halo Finder
Comparison Project \citep{MAD} lists 29 different codes and roughly
divides them into two distinct groups:
\begin{enumerate}
\item Percolation algorithms, for which particles are linked together
  if closer to each other than some specified linking length. The
  typical example is the algorithm ``\emph{friends-of-friends}''
  (thereafter FOF) \citep{FOF}.
\item Segmentation algorithms, for which space is segmented into
  separate regions around local peaks of the density field. Particles
  within these regions are then collected and assigned to the same
  halo or sub-halo. The typical example is the ``\emph{Spherical
  Overdensity}'' method (thereafter SOD) \citep{SO}.
\end{enumerate}
The outer boundary of the haloes are defined in both method by a
density iso-surface, whose exact value determines the properties of
the resulting halo statistics.  Halo catalogues derived from FOF and
SOD and their corresponding merger trees have been studied quite
extensively in the literature \citep[see e.g.][]{SUSSING_HALOFINDER}.

\subsection{The PHEW halo finder}

In this paper, we extend these earlier studies to the \phew\ halo
finder \citep{PHEW} developed specifically for the \ramses\ code
\citep{ramses}.  The \phew\ algorithm belongs to the category of
segmentation methods.  Particle masses are first deposited to the AMR
grid using the ``\emph{cloud-in-cell}'' technique. All density maxima
are then marked as potential sites for a \emph{clump}. Clumps are what
we call any structure, haloes and sub-haloes, in contexts where we
don't need to differentiate between them. The volume is then segmented
into peak patches by assigning each cell of the grid to the closest
density maximum in the direction of the steepest density gradient.

This segmentation method provides well defined regions separated by
density saddle surfaces. The minimum density in the saddle surface
between two adjacent peaks marks the saddle point between the two
peaks.  This well known method is often called `\emph{watershed
segmentation}`. In order to define proper halo boundaries, \phew\ uses
an outer density isosurface, like most methods described
above. Subhaloes, on the other hand, are just the ensemble of all
peak patches within the halo boundaries. This allows to identify
haloes and sub-haloes without the assumption of spherical symmetry,
unlike other popular methods such as SOD.

Subhaloes can be organised into a hierarchy of sub-structures based
on the same steepest gradient technique, for which individual clumps
can be assigned to the closest densest peak.  After this first pass,
only a few sub-haloes survived the merging process, which is then
repeated a second time, assigning these surviving sub-haloes to their
densest neighbours.  Ultimately, all sub-haloes will be collected into
a single peak that corresponds to the main halo. Each pass defines a
level in the hierarchy of sub-haloes. More details can be found in the
original \phew\ paper \citep{PHEW}.  As a consequence, a halo can have
a number of sub-haloes, each one of them containing subsub-haloes and
so on.  This well defined hierarchy is a very important feature for us
to uniquely assign particles to haloes and sub-haloes based on a
binding energy criterion.
 
There are four parameters that \phew\ requires a user to choose in
order to identify clumps and haloes.  Firstly, a ``relevance
threshold'' needs to be defined.  If for any given peak patch the
ratio of the peak's density to the maximal density of the entire
saddle surface of the respective peak patch is smaller than the chosen
relevance threshold, then the peak patch is considered to be noise,
not a genuine structure.  The peak patch is then merged into a
neighbour.  Secondly, a density threshold determines the minimal
density a cell needs to have to be part of any peak patch.  Thirdly, a
``saddle threshold'' defines the maximal density for a saddle surface
between two peak patches for the two patches to be considered parts of
two different haloes.  If the saddle surface density is above the
threshold, then the peak patches will be parts of the same halo (but
different sub-haloes within the host halo).  Finally, a mass threshold
determines the minimal mass a peak patch needs to have to be kept.  We
list the parameters that we used throughout this paper in Table~\ref{tab:phew-parameters}.

\begin{table}
\centering
\caption{Parameters used for the \phew\ clump finder throughout this
  paper.  The numerical values given can be directly used in the
  namelist file that \ramses\ uses to read in runtime parameters.
  Here $\bar \rho$ here is the mean background density and $m_p$ is
  the particle mass.}
\label{tab:phew-parameters}
\begin{tabular}[c]{l l l}
  parameter				&	value		& units \\
  \hline
  relevance threshold			&	3		& 1		\\
  density threshold			& 80			& $\bar \rho$ 	\\
  saddle threshold			& 200			& $\bar \rho$	\\
  mass threshold			& 10			& $m_p$		\\
  \hline
\end{tabular}
\end{table}
 
%Structure identified in this manner needs to be checked for being true condensations as opposed to arising from Poisson noise.	
%Such ``\emph{noise}'' peak patches are detected by computing the ratio of the density of each peak to the highest density of any cell of the peak patch that borders on another peak patch.
%If the ratio is sufficiently low, the patch is deemed to be noise and merged into the peak patch with whom it shares the aforementioned cell on its border with the highest density.
%Once the noise is identified and removed, the remaining structure consists only of peak patches, essentially clumps of particles, which satisfy the relevance condition. 
%These clumps represent the structure on the lowest scale.
%A large halo for example, which can very roughly be described as ``a large clump'' in a first approximation, would be decomposed into many small clumps at this point.
%The low level structure needs to be merged into composite clumps to form large scale haloes. 
%This merging is done iteratively and in doing so the hierarchy of substructure is established.
%A consequence, however, is that by construction, substructure will always have at least one neighbouring parent structure, which will have influence on the definition of when a particle is bound to the substructure.
	
%==========================================================
%\subsection{Particle Unbinding}\label{chap:unbinding}
%==========================================================

\subsection{Particle unbinding}

We now describe how we assign each dark matter particle to a given
sub-halo, a process that has not been implemented so far in the
\phew\ code.  For this, we follow a physically motivated criterion,
quite common in the halo finding literature, based on the binding
energy of the particle \citep[e.g.][]{AHF, subfind, skid}.  If a
particle is not bound to the first sub-halo of the hierarchy, it is
then passed recursively to the next sub-halo in the hierarchy, where
the binding energy is checked again and so on.  If the particle is not
bound to any sub-halo, it is assigned to the main halo.

\begin{figure}
  \centering
  \fbox{\includegraphics[width=.95\linewidth, keepaspectratio]{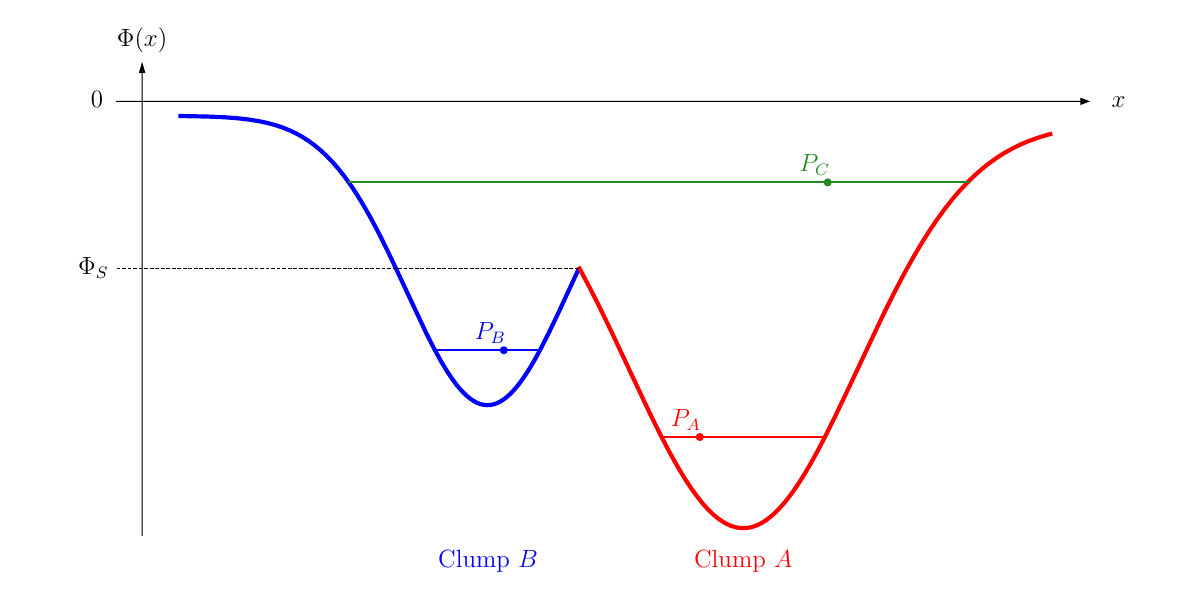}}%
  \caption{Simple sketch of the gravitational potential of a halo that
    consists of two clumps, $A$ and $B$.  The position of the
    horizontal lines marks the energies of three example particles,
    $P_A$, $P_B$, and $P_C$, while the length of the lines shows the
    spatial extent of the orbits.  We call particles like $P_A$ and
    $P_B$ ``strictly bound'', as their predicted orbit boundaries
    don't allow them to escape from the clump they are assigned to.
    Particle $P_C$ however, although energetically bound to clump $A$,
    can wander off deep into clump $B$, and for that reason is called
    ``loosely bound''. To discriminate between these two types of
    particles, we use the potential of the saddle point between clump
    $A$ and $B$ marked as $\Phi_S$.
  }%
  \label{fig:potentials}
  %	\endminipage\hspace*{\fill} 
\end{figure}

In the previous hierarchical unbinding process, the key component is
the criterion adopted for deciding whether a particle is bound to a
sub-halo or not.  Traditionally, this is done using the {\it static}
gravitational potential, since we are dealing with a single time step
and we have to assume that the N-body system is stationary.  In this
case, a particle at position ${\bf r}$ is considered as unbound if its
velocity ${\bf v}$ exceeds the escape velocity given by
\begin{equation}
v_{\rm esc} = \sqrt{- 2\Phi(\mathbf{r})} 
\label{eq:boundv}
\end{equation}
More precisely, this means that the particle will be able to travel to
infinity where the potential goes to  zero. If the velocity is smaller
than the escape  velocity, the particle will follow a  bound orbit and
come back to its current location.  Note that this orbit can leave the
boundaries of the  sub-halo.  The particle will stay for  some time in
the sub-halo,  but can visit at  a later time a  neighbouring sub-halo
and then come back along the  same bound orbit.  This kind of particle
does not {\it exclusively} belong  to its original sub-halo. It should
be in fact assigned to the parent sub-halo in the hierarchy.  In order
to identify particles as more  strictly bound, we re-define the escape
velocity using the potential of the closest saddle point $\Phi_S$.
\begin{equation}
v_{\rm esc} = \sqrt{- 2(\Phi(\mathbf{r})-\Phi_S)} 
\label{eq:boundv_corr}
\end{equation}
This new  escape velocity is  smaller than the previous  one, allowing
more particle to  exceed it and leak out of  the current sub-halo into
neighbouring ones.  In what  follows, we will  use these two unbinding
criteria, calling the first method the ``loosely bound'' criterion and
calling   the   second   one   the   ``strictly   bound''   criterion.
Figure~\ref{fig:potentials} illustrates  the difference  between these
two  criteria.   The  gravitational  potential  of   two  neighbouring
sub-haloes labelled A  and B is represented.  We show  an example of a
strictly bound particle in each sub-halo,  and an example of a loosely
bound particle that can wander from one sub-halo to the other one.  If
one uses the first binding criterion, this loosely bound particle
will be  assigned to sub-halo A,  because this is where  it is located at
the present time.  If one uses the second, stricter binding criterion,
then the loosely bound particle will be assigned to the parent halo, but
not to sub-halo A nor B.

%A typical definition of a 'bound particle' can be obtained by considering an isolated clump in a time-independent scenario, where energy is conserved, a particle $i$ is considered to be bound if its velocity is smaller than the escape velocity:
%\begin{equation}
%	v_i < \sqrt{- 2 \cdot \Phi(\mathbf{r}_i)}  \quad \Leftrightarrow \quad \text{ particle is bound }\label{eq:boundv}
%\end{equation}

%where $\Phi$ is the gravitational potential, $\mathbf{r}_i$ is the particle's position and $v_i = ||\mathbf{v}_i||$ is the magnitude of the particle's velocity of each particle $i$, both given in the frame of reference of the centre of mass of the clump.
%An approximation for the potential $\Phi$ can be found by assuming spherical symmetry and solving the Poisson equation in the centre of mass frame of the clump.
%\begin{equation}
%	\Delta \Phi = \frac{1}{r^2}\frac{\del}{\del r} \left(r^2 \frac{\del}{\del r} \Phi \right) = 4 \pi G \rho
%\end{equation}

%One problem with condition \ref{eq:boundv} is that it assumes that clumps will be isolated, which by construction of the clump finder \phew\ sub-haloes will never be.
%The issues that arise from this fact can be understood by considering the boundaries of a particle's trajectory.
%These boundaries in a given potential $\Phi$ can be estimated using the conservation of energy:
%
%\begin{align}
%	E/m_p = \frac{1}{2} v^2 + \Phi = const.
%\end{align}
%

When computing the velocity of the particle, it is important to use
the velocity {\it relative} to the velocity of the sub-halo centre of
mass (also called the bulk velocity of the sub-halo).  Because of this
requirement, the unbinding process has to be performed iteratively,
since removing a particle that is unbound requires to recompute the
centre of mass velocity.  Let us finally repeat that the particle
unbinding is performed recursively, following the sub-halo hierarchy
from the bottom up.  Starting with the lowest (finest) level of
sub-haloes, unbound particles are assigned to the higher (coarser)
level of parent sub-halo for unbinding, and so on following the
structure hierarchy.  Particles that are unbound from all sub-haloes
are collected into the main halo, marking the end of the hierarchical
unbinding process.

    %====================================================================
\section{Merger Trees: Basic Principles}\label{chap:making_trees}
%====================================================================

In this work, we adopt the terminology set by the ``Sussing Merger
Tree Comparison Project'' \citep{SUSSING_COMPARISON,
  SUSSING_CONVERGENCE, SUSSING_HALOFINDER,leeSussingMergerTrees2014a}.
For sake of clarity, we repeat here some important definitions:

\begin{itemize}

\item For two snapshots at different times, a halo from the first one
  (i.e. higher redshift) is always referred to using the capital
  letter $A$ and a halo from the second one (i.e. lower redshift)
  using $B$.

\item Recursively, $A$ itself and progenitors of $A$ are all
  \emph{progenitors} of $B$.  When it is necessary to distinguish $A$
  from earlier progenitors, the term \emph{direct progenitor} will be
  used.

\item Recursively, $B$ itself and descendants of $B$ are all
  \emph{descendants} of $A$.  When it is necessary to distinguish $B$
  from later descendants, the term \emph{direct descendant} will be
  used.

\item In this work, we restrict ourselves to merger trees for which
  there is \emph{precisely one direct descendant for every halo}.

\item When there are multiple direct progenitors, it is required that
  one of these is identified as the \emph{main progenitor}.

\item The \emph{main branch} of a halo is a complete list of main
  progenitors tracing back along its cosmic history.

\end{itemize}
We finally define an important convention we use here: when no
distinction between sub-haloes and main haloes is necessary, they are
collectively referred to as \emph{clumps}.

\subsection{Linking Clumps Across Snapshots}

\begin{figure}
  \includegraphics[width=\linewidth]{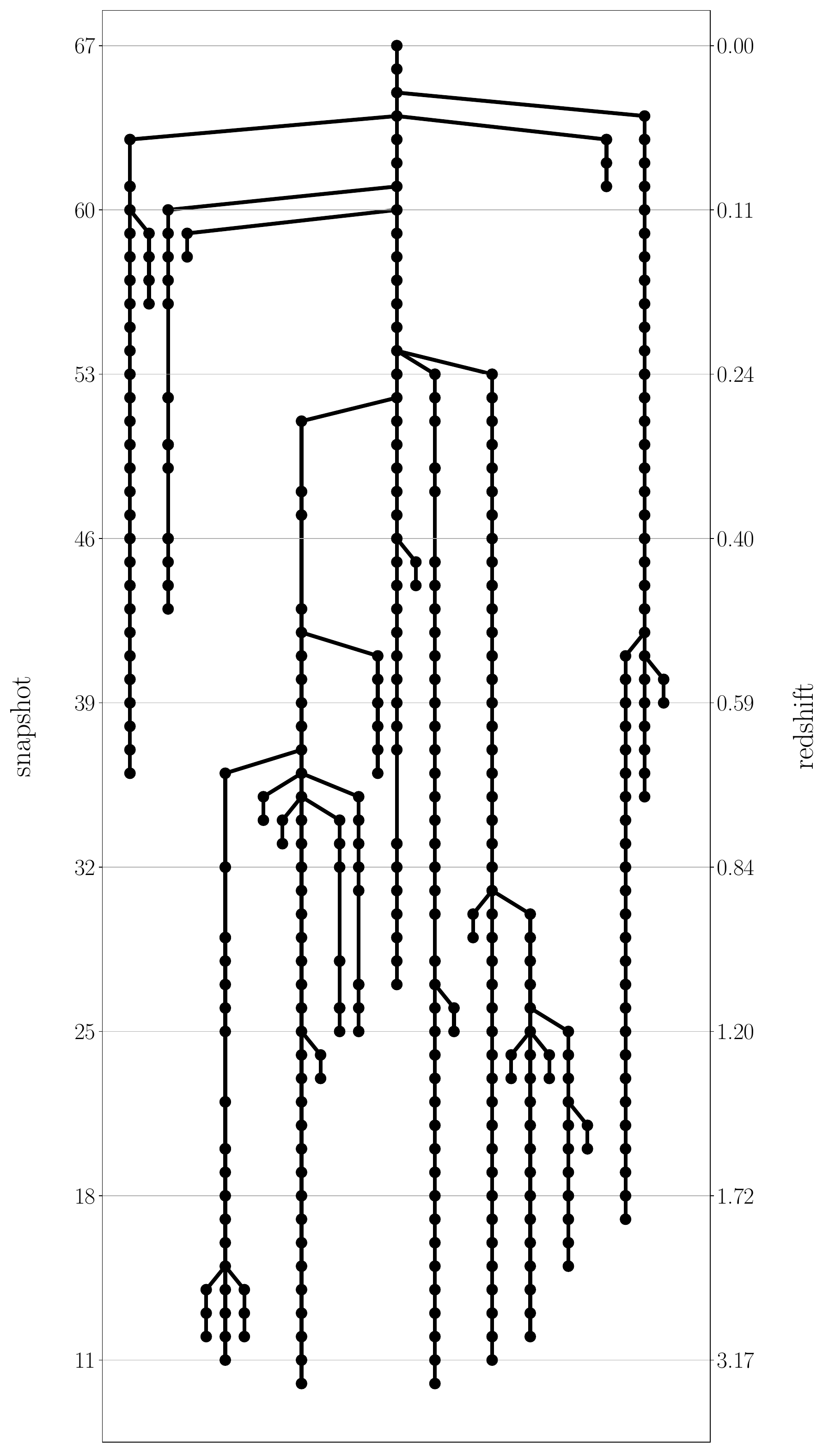}%
  \caption{The merger tree of a main halo at redshift zero as found by
    \texttt{ACACIA}.  This tree was extracted from a low resolution
    cosmological DMO simulation containing $64^3$ particles for
    illustrative purposes. Using higher resolutions quickly leads to
    hundreds and thousands of branches, resulting in a rather messy
    plot. On the $y$ axis, the snapshot numbers and their
    corresponding redshifts are given.  The $x$ axis has no physical
    meaning.  Each dot represents a clump identified at the given
    snapshot.  Two dots connected by a vertical line represent clumps
    that have been linked as main progenitor and main descendant.
    Diagonal lines depict merging events. In this tree, a few links
    between two dots are larger than others.  These are cases where
    clumps merge temporarily, but then re-emerge later as separate
    clumps (see the example shown in Figure~\ref{fig:jumper-demo}).
    We discuss these cases and how they are dealt with in
    Section~\ref{sect:jumpers}. }
  \label{fig:mergertree}
\end{figure}

The aim of a merger tree code is to link haloes from an earlier
snapshot to haloes in the consecutive snapshot, i.e. to find all the
descendants of the haloes in the earlier snapshot. If we do this
successfully each snapshot, then we can follow the formation history of
haloes throughout the simulation. In particular, this will enable us
to track the mass growth of haloes as well as the merging of different
sub-haloes during the course of the simulation.
To illustrate the idea, a merger tree of a main halo generated by
\texttt{ACACIA} during a simulation is shown in Figure~\ref{fig:mergertree}.
In this particular case, we were able to track the formation history of the
main halo down to redshifts $z > 3$. By linking progenitors and descendants
throughout the simulation many branches of the tree are revealed.
Each branch represents a clump that eventually merged into the main halo
which we chose as the root of the tree at redshift zero.

Merger events occur when a clump, identified as such in a previous
snapshot, disappears from the list of clumps in the next snapshot. In
the case of a halo, it usually first becomes the sub-halo of another
halo and can be followed as such in many subsequent snapshots. After
some time, this sub-halo can merge into another sub-halo or dissolve
completely due to numerical over-merging. In both cases, the sub-halo
disappears completely from the clump catalogue.

A straightforward method to link progenitors with descendants in two
consecutive snapshots is to trace individual particles using their
unique particle ID. All merger tree codes use this simple technique
\citep[][]{ConsistentTrees, LHaloTree, D_Trees,
  knebeImpactBaryonicPhysics2010, tweedBuildingMergerTrees2009,
  elahiClimbingHaloMerger2019a, jungEffectsLargescaleEnvironment2014,
  rodriguez-gomezMergerRateGalaxies2015} with the notable exception of
the code \texttt{JMERGE} described and tested in
\cite{SUSSING_COMPARISON}.

Linking a progenitor to a descendant means checking how many particles
of the progenitor halo or sub-halo end up in the descendant halo or
sub-halo.  Naturally, these tracer particles may end up in multiple
clumps, giving multiple descendant candidates for a progenitor.  In
such cases, the most promising descendant candidate will be called the
\emph{main descendant}.  To find a main progenitor and a main
descendant, a merit function $\mathcal{M}$ has to be defined, which is
to be maximised or minimised, depending on its definition.  An
overview of the merit functions that are used in other merger tree
algorithms is given in Table~1 of \cite{SUSSING_COMPARISON}.  The
merit function used in our implementation is given in
Equation~\ref{eq:merit}.

Sometimes, unfortunately, linking progenitors to descendants is not as
straightforward as described so far. We now discuss two circumstances
where special care must be taken to define robust links between
different snapshots: fragmentation events and temporary merger events.

%Because galaxies form inside the potential well dark matter haloes,
%knowledge of how many merging events a halo underwent during its
%lifetime is crucial for accurate mock galaxy catalogues.  After a
%merging event, the galaxy of the smaller halo that has been
%``swallowed up'' by a bigger one has no reason to simply vanish
%without a trace.  The ``swallowed up'' halo might become a sub-halo,
%or, if it is small enough or after some time, it might not be
%detectable as substructure in the simulation any more.  Galaxies of
%haloes that dissolve in this manner are referred to as ``\emph{orphan
%galaxies}'' \citep[e.g.][]{subfind}.

\begin{figure}
  \centering
  \includegraphics[width=.9\linewidth]{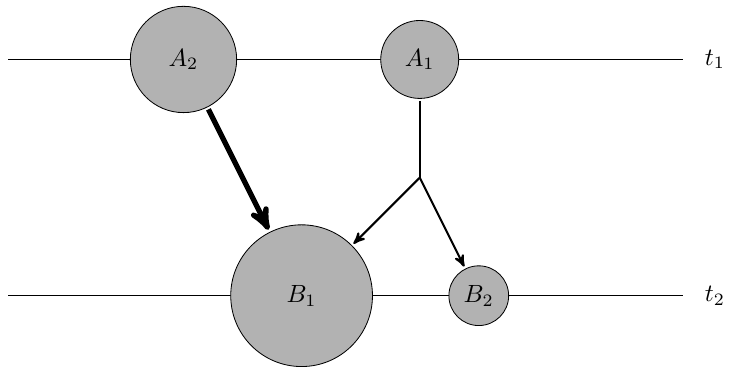}
  \caption{Illustration of a fragmentation event, where a
  	progenitor $A_1$ at time $t_1$ is partially merged together
  	with a second progenitor $A_2$ into a descendant $B_1$ at
  	time $t_2 > t_1$, while some fragmented part of $A_1$, $B_2$,
  	evaded the merging.
	}
  \label{fig:fracture}
\end{figure}

\subsection{Fragmentation Events}
\label{sect:frag}

%See https://arxiv.org/pdf/2010.03567.pdf Fig. 34
In our current approach, each progenitor can have only one
descendant\footnote{Note that \cite{springelSimulatingCosmicStructure2021a}
proposed another approach that allows explicitly fragmentation
events to be included in the formation history analysis, where they use
merger \emph{graphs} rather than merger trees.}. We therefore need
to pick only one descendant within a possibly large ranked list of
descendant candidates, that all contain particles coming from the
progenitor.

Normally, this choice is performed according to the ranking provided
by the merit function, where the main descendant is ranked number 1.
Problems arise for example when the progenitor $A_1$ is not the main
progenitor of its main descendant $B_1$, but also has fragmented into
another viable descendant candidate $B_2$.  This situation is
schematically shown in Figure~\ref{fig:fracture}.

Relying only on the merit function \eqref{eq:merit}, progenitor $A_1$
will seem to have merged with $A_2$, the direct progenitor of $B_1$,
in order to form $B_1$.  The other fragment, $B_2$, will be treated as
a newly formed clump and the entire formation history of $B_2$ would
be lost. In order to preserve this history, we choose to prioritize the
link from $A_1$ to $B_2$ over of merging progenitor $A_1$ into $B_1$.

It is simpler to deal with this case directly in the algorithm
than via the merit function. The resulting logic can be summarized as
follows: If $A_1$ is not the main progenitor of its main descendant
$B_2$, then we don't merge it into $B_2$ but we link it instead with
the first secondary descendants that considers $A_1$ as its
main progenitor.

%This way, we give priority to $A_1$ for being a progenitor to some descendant which is
%not its main descendant over merging it into its main descendant.  In
%other words: being a main progenitor will have more weight on the tree
%building scheme than being a main descendant.

\subsection{Temporary Merger Events}
\label{sect:jumpers}

\begin{figure*}
    {
        \setlength\tabcolsep{0em}
    	\centering
    	\begin{tabular}{p{5.3cm}p{5.3cm}p{5.3cm}}
            \centering
%    		\hline
    		%		 
    		\fbox{\includegraphics[width = 5.1cm]{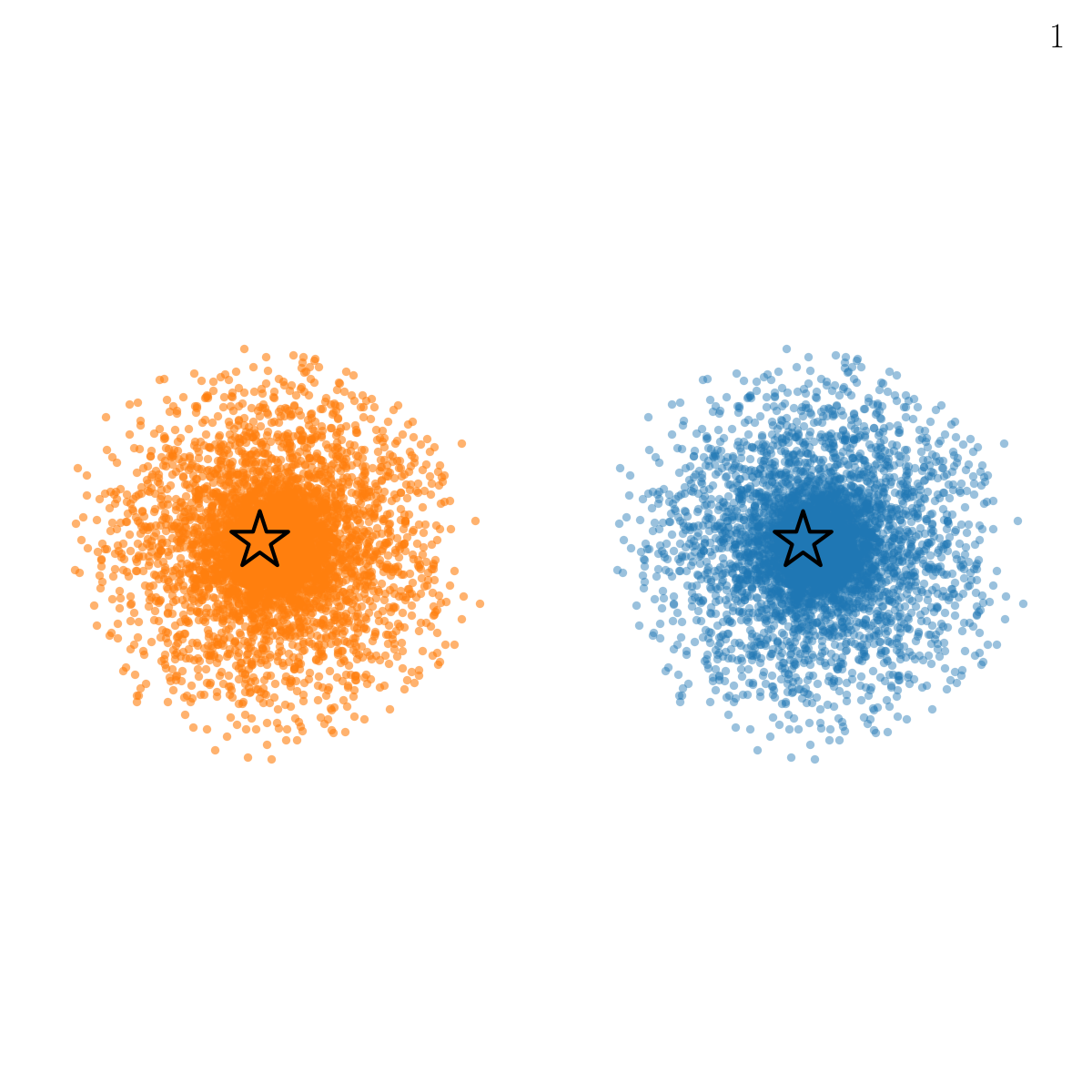}}	& 
    		\fbox{\includegraphics[width = 5.1cm]{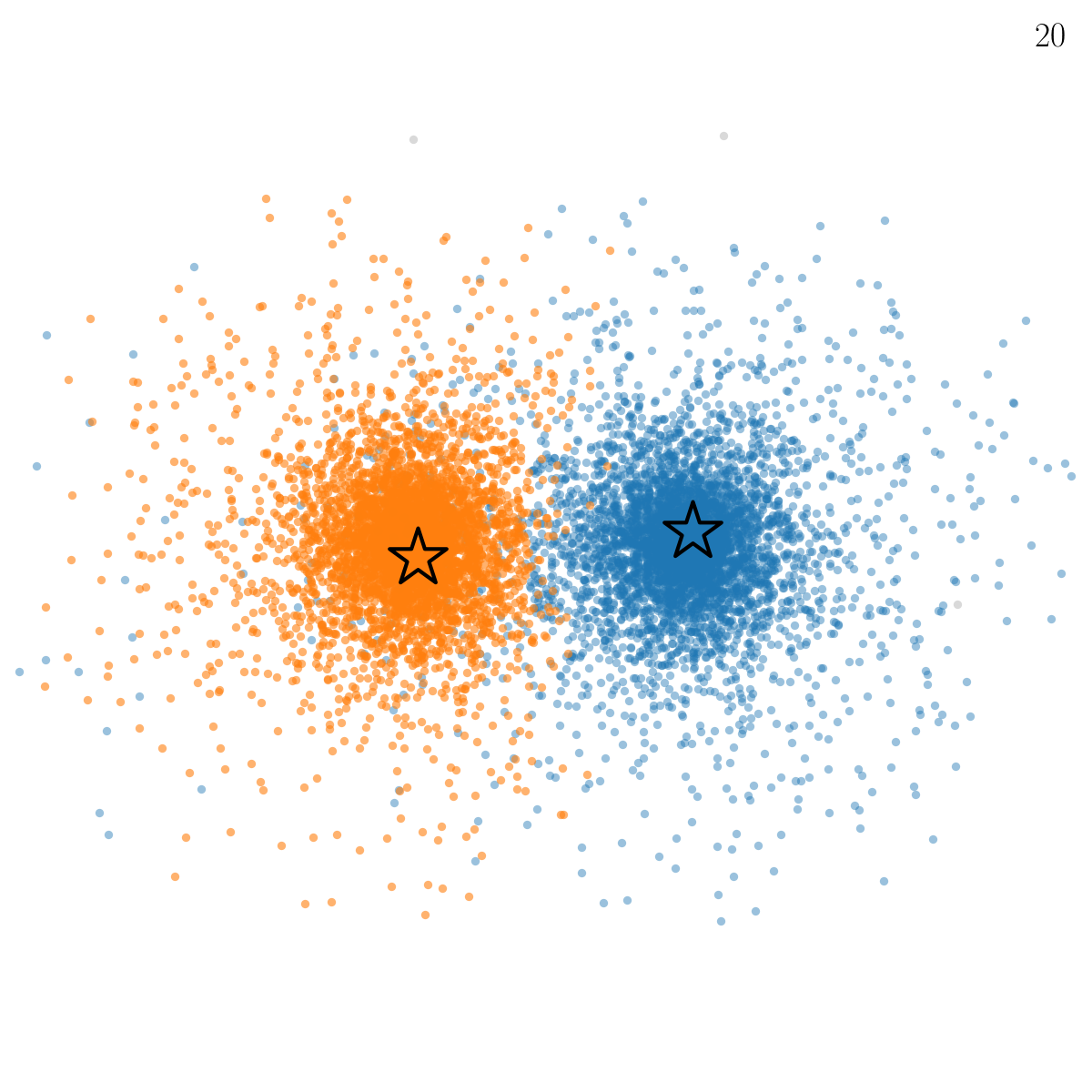}}	& 
    		\fbox{\includegraphics[width = 5.1cm]{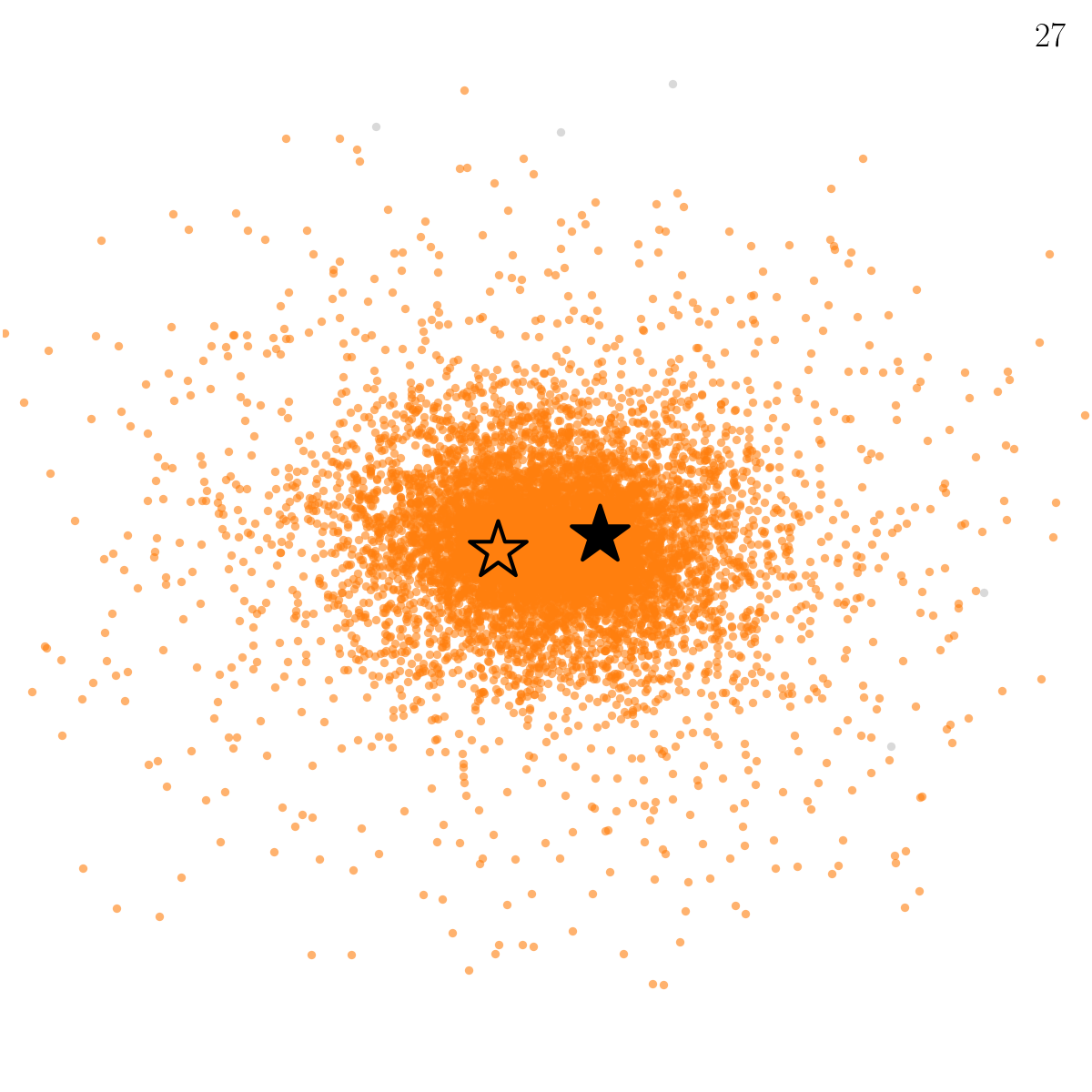}}  \\%[-0.5em]
    		\fbox{\includegraphics[width = 5.1cm]{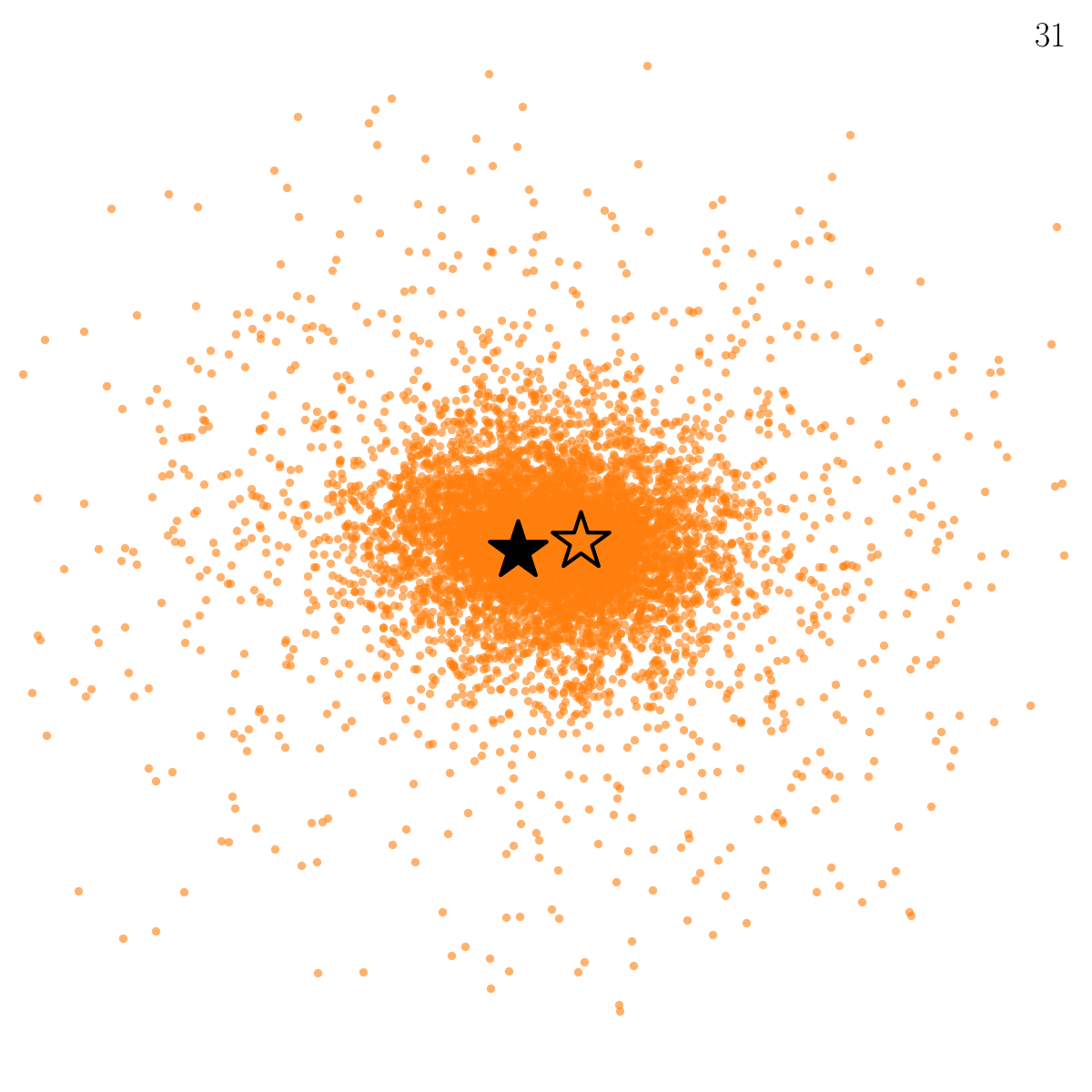}}	& 
    		\fbox{\includegraphics[width = 5.1cm]{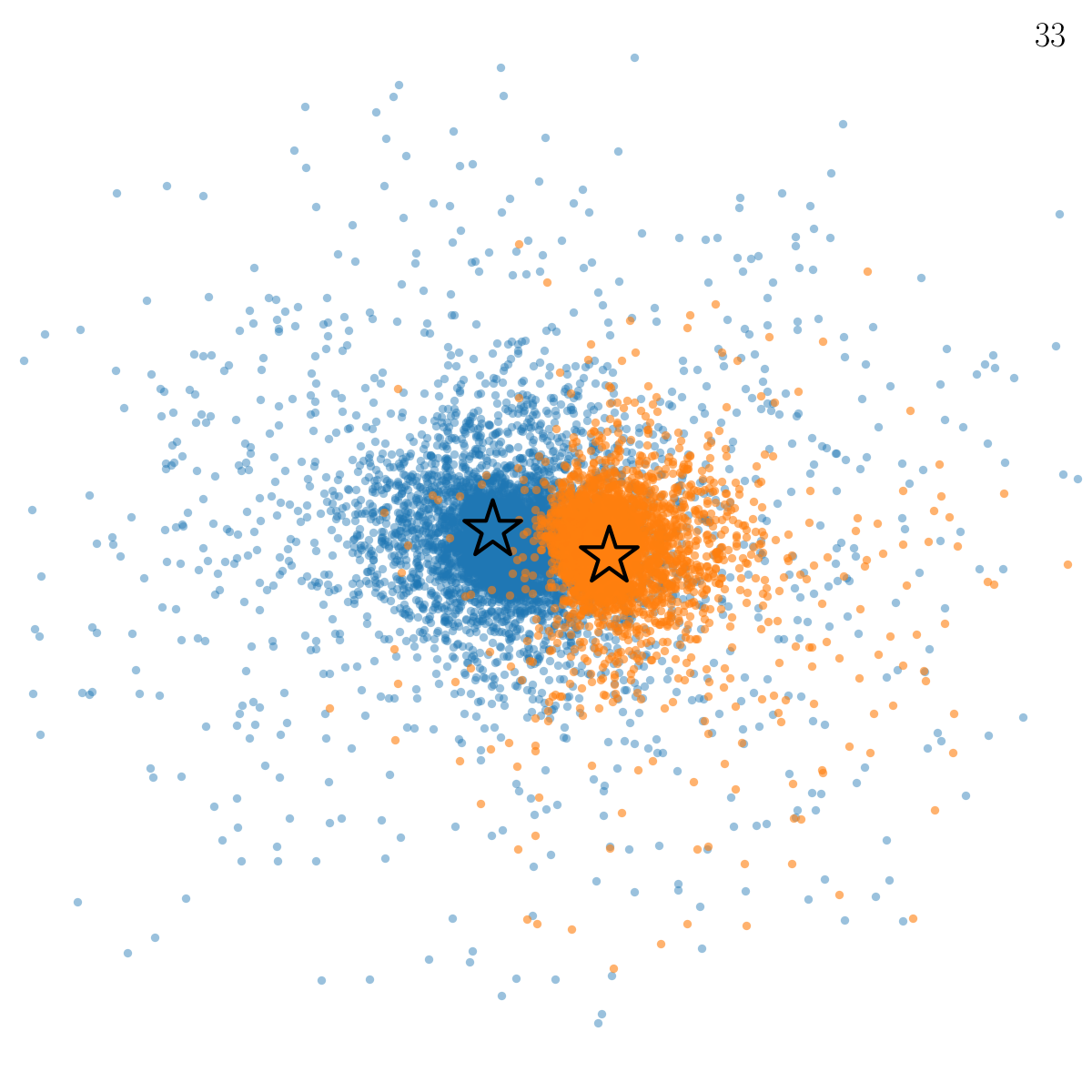}}	& 
            \fbox{\includegraphics[width = 5.1cm]{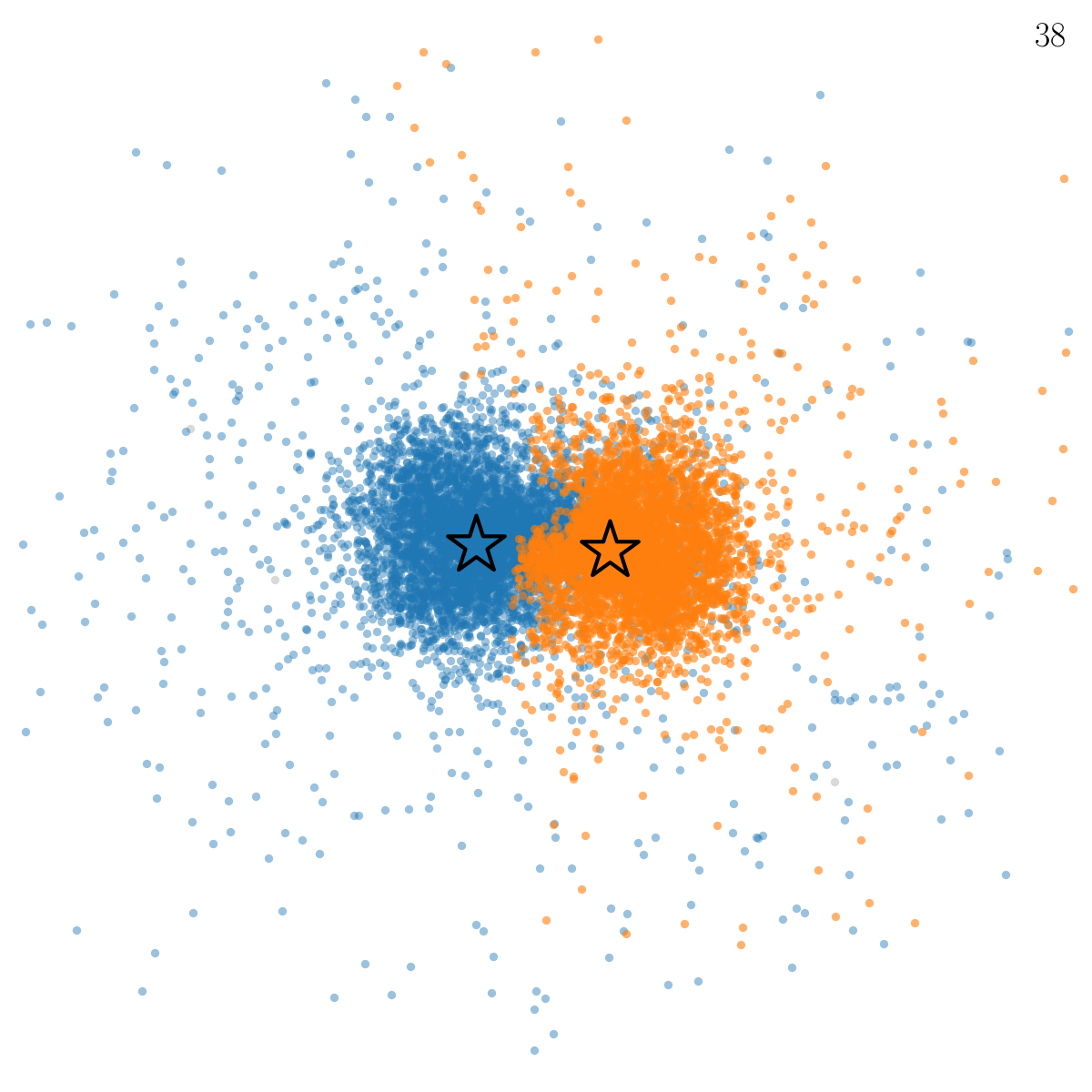}}	\\%[-0.5em] 
    	\end{tabular}	
%    	\begin{tabular}{p{5.3cm}p{5.3cm}p{5.3cm}}
%            \centering
%%    		\hline
%    		%		 
%    		\fbox{\includegraphics[width = 5.1cm]{images/jumper-demo/particleplot_00001.png}}	& 
%    		\fbox{\includegraphics[width = 5.1cm]{images/jumper-demo/particleplot_00018.png}}	& 
%    		\fbox{\includegraphics[width = 5.1cm]{images/jumper-demo/particleplot_00024.png}}  \\%[-0.5em]
%    		%
%    		%
%    		\fbox{\includegraphics[width = 5.1cm]{images/jumper-demo/particleplot_00030.png}}	& 
%    		\fbox{\includegraphics[width = 5.1cm]{images/jumper-demo/particleplot_00039.png}}	& 
%            \fbox{\includegraphics[width = 5.1cm]{images/jumper-demo/particleplot_00042.png}}	\\%[-0.5em] 
%    	\end{tabular}
    }
	\caption{\label{fig:jumper-demo} 
        Illustration of how haloes can seemingly merge into another one and re-appear a few snapshots later.
        The blue and orange particles are two initially distinct haloes that pass through each other.
        The galaxies assigned to them are marked by a star with the same colour as the particles.
        Fully black stars mark orphan galaxies, which have lost their unique host (sub)halo.
        The number in the upper right corner of each plot is the snapshot number that is depicted.
        In snapshots 27-32, the halo-finding algorithm doesn't identify both haloes as distinct objects.
        However by tracing the blue halo's orphan galaxy it was possible to link the halo in snapshot 33 all the way back to snapshot 26. \hspace{\textwidth}
        The initial conditions were created using \texttt{DICE} \citep{DICE}.
%        Both haloes are identical with mass of $5\cdot 10^{10}\msol$, each containing 5000 particles and following a NFW mass profile.
%        The plotted region corresponds to $400$ kpc on each side.
        }
\end{figure*}

When a sub-halo travels towards the core of its parent halo, it will
merge with the central clump and disappear from the sub-halo lists.  It
can however re-emerge at a later snapshot and will be added back
to the list as a newly born halo. Such a scenario is shown in
Figure~\ref{fig:jumper-demo}.  Indeed, when this occurs, the merger
tree code will deem the sub-halo to have merged into the main halo, and
will likely find no progenitor for the re-emerged sub-halo, thus
treating it as newly formed.

This is a problematic case because we lose track of the growth history
of the sub-halo, regardless of its size, and massive clumps may be
found to just appear out of nowhere in the simulation.  This is a well
known problem for configuration-space halo finders
\citep{onionsSubhaloesGoingNotts2012}, and phase-space halo finders
like \texttt{ROCKSTAR} \citep{behrooziRockstarPhaseSpaceTemporal2013}
have been developed precisely to alleviate this issue.  While they
typically perform better than configuration-space halo finders,
\cite{SUSSING_COMPARISON} found that phase-space halo finders aren't
infallible in recovering all such missing haloes, and strongly
recommend checking for links between progenitors and descendants in
non-consecutive snapshots as well.

As our merger tree code works on the fly, future snapshots will not be
available at the time of the merger tree analysis, so it will be
necessary to check for progenitors of a descendant across multiple
snapshots.  This can be achieved by keeping track of the most bound
particles of each clump when it is merged into some other clump.
These tracer particles are also used to track {\it orphan galaxies}.
\footnote{In the context of SAM, orphan galaxies are galaxies born at
the center of dark matter haloes that merged later into bigger haloes
and eventually dissolved due to over-merging. As a consequence, these
galaxies don't have a parent halo or sub-halo anymore.}  For this
reason, we call these tracer particles ``orphan particles'', and
progenitor-descendant links over non-adjacent snapshots ``jumpers''.

These jumper links between different haloes widely separated in time
are less reliable than proper links between progenitors and
descendants from adjacent snapshots.  As we will discuss
in Section~\ref{chap:testing-nmb}, the quality of the merger trees
increases with the number of tracer particles used.  Using jumpers
corresponds to using only one tracer particle over a large time interval,
much larger than the one between two adjacent snapshots.

For this reason, priority is given to direct progenitor candidates in
adjacent snapshots.  Only if no direct progenitor candidates have been
found for some descendant, then progenitor candidates from
non-adjacent snapshots are searched for.  Because these progenitors
from non-adjacent snapshots are only tracked by one single particle,
we don't use the merit function to rank them.  Instead, we find the
orphan particle within the descendant clump which is the most tightly
bound.

In conclusion, although not ideal, using jumpers remains a necessity
to track these temporary merger events.  As a bonus, it allows us to
track orphan galaxies as will be discussed in
Section~\ref{chap:mock_catalogues}.

%============================================
\section{Details Of Our New Algorithm}\label{chap:my_code}
%============================================

\begin{figure*}
	\includegraphics[width=\textwidth]{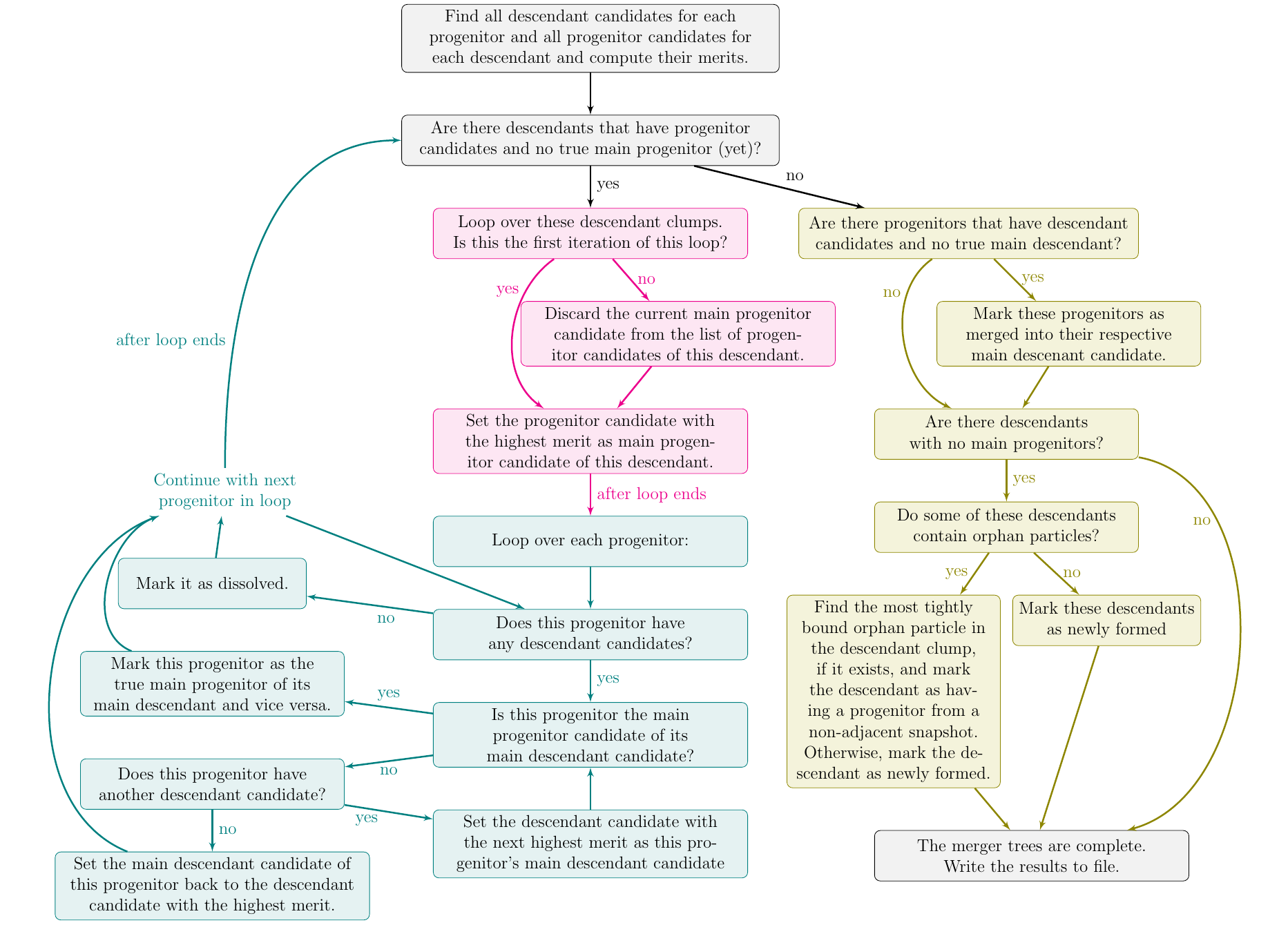}%
	\caption{\label{fig:flowchart}
    	Flowchart of the tree making algorithm. For more details, please refer to Section~\ref{chap:my_code} in the main text.
  }
\end{figure*}

For clarity, the algorithm that we describe in what follows
is also shown in a flow diagram in Figure \ref{fig:flowchart}.
The first step of our algorithm is to identify plausible progenitor
candidates for descendant clumps, as well as descendant candidates for progenitor clumps. We achieve this by tracking tracer particles
across simulation snapshots. For any given snapshot, the tracer particles
of each clump are selected within the list of all particles belonging to
the clump, ranked from most bound to least bound. Indeed, the most bound particles are expected to remain well within the clump boundary between
two snapshots. The main parameter of our method is the maximum number of
these tracer particle used per progenitor clump, called $n_{\rm mb}$. The minimum number of tracer particles is obviously equal to $n_{\rm min}$, the minimum mass threshold in units of particle masses adopted by the \phew\ clump finder.

So for every clump in the current snapshot, the $n_{\rm mb}$
most bound tracer particles are found and written to file. In the following
output step, those files are read in and the tracer particles
are used to determine which clumps of the previous snapshot are the
progenitor candidates of the clumps of the current snapshot.
For each progenitor clump, we compile a list of descendant clump candidates by finding all descendant clumps which contain tracer particles of said progenitor. Conversely, we also compile a list of progenitor candidates for each descendant clump by finding all tracer particles amongst the descendant clump's particles and by noting which progenitor clump they are tracing.

As explained earlier, in order to generate the merger trees the \textit{main} progenitor of each descendant and the \textit{main} descendant of each progenitor need to be found. Commonly at this point however, multiple progenitor candidates have been found for every descendant, as well as multiple descendant candidates for each progenitor. Therefore, we somehow need to select the ``best'' candidate amongst those with the aim of generating reliable merger trees. To be able to select the ``best'', we first quantify how ``good'' a candidate is by assigning a merit to each descendant candidate of every progenitor, as well as every progenitor candidate of each descendant. We define the merit as follows:

Let $\mathcal{M}_{\rm pd}(A,B_i)$ be the merit function to be
maximised for a list of descendant candidates $B_i$ of a progenitor
$A$. Let $n_{\rm mb}$ be the total number of particles of progenitor
$A$ that are being traced. Note that $n_{\rm mb}$ can be smaller than the
total number of particles in clump $A$.  A straightforward ansatz for
the merit function would be to based on the fraction of particle traced from the
progenitor to the descendant candidate:
\begin{equation}
\mathcal{M}_{\rm pd}(A,B_i) \propto \frac{n_{A \cap B_i}}{n_{\rm mb}}
\end{equation}
where $n_{A \cap B_i}$ is the number of tracer particles of $A$ found
in $B$.  Similarly, we define $\mathcal{M}_{\rm dp}(A_i,B)$ as the
merit function to be maximised for a list of progenitor candidates
$A_i$ of a descendant $B$. Another straightforward ansatz would be
based on the fraction of particle traced from the progenitor
candidates to the descendant:
\begin{equation}
\mathcal{M}_{\rm dp}(A_i,B) \propto \frac{n_{A_i \cap B}}{n_B}
\end{equation}
where $n_B$ is the total number of particles in the descendant $B$.
In these two merit functions, $n_{\rm mb}$ and $n_B$ are just
normalizing factors. They are independent of the properties of the
candidate and hence won't affect the selection process.  We can
therefore define a generic merit function as
\begin{equation}
\mathcal{M}(A,B) \propto n_{A \cap B}
\end{equation}

The \phew\ clump finder in \ramses\ identifies the main halo as the
clump with the highest density maximum.  During a major merger event,
the halo will have two clumps with similar masses and comparable
maximum densities.  It is then quite common that small variations in
the value of the density maxima will cause the identification of the
main halo to jump between these two clumps.  Indeed, the particle
unbinding algorithm will identify particles that are not bound to the
sub-halo and pass them on to the main halo, modifying the resulting
mass for the two clumps.  This effect is particularly strong if one
uses the {\it strictly bound} definition for the particle
assignment. As a consequence, because the identification of the main
halo varies, strong mass oscillations are expected. To counter this
spurious effect, we modify the merit function to preferentially select candidates with similar masses:
\begin{equation}
\mathcal{M}(A,B) = \frac{n_{A \cap B}}{n_{\rm max} - n_{\rm min}} \label{eq:merit}
\end{equation}
where $n_{\rm max}=\max(n_A,n_B)$ and $n_{\rm min}=\min(n_A,n_B)$.  An
overview of other merit functions used in the literature is given in
Table~1 of \cite{SUSSING_COMPARISON}.

With the merit function defined, let's return to the tree making algorithm. We left off with the commonly encountered situation in which we have identified multiple progenitor candidates for descendants and vice versa, and are now
looking to find a matching progenitor-descendant pair, where the \textit{main} progenitor candidate of a descendant has precisely this descendant as its \textit{main} descendant candidate. This search is performed iteratively.
%A main progenitor-descendant pair is
%established when the main progenitor of a descendant is the main
%descendant of said progenitor.
%At every iteration, we check all the descendant candidates that haven't found
%their match yet. For each of these descendants, we check all of their progenitor
%candidates
At every iteration, we first look at all progenitors that haven't found their respective match yet, and check all of their descendant candidates for a match. The loop over descendant candidates for a given progenitor is performed in order of decreasing merit of the descendant candidates, and the checks are  stopped either when a progenitor has found a match, or has run out of candidates. (In case a progenitor has no descendant candidates at all, we consider it as dissolved.)
Once all the progenitors have been checked, the second step of the iteration begins. We now look at all descendants that haven't found their respective match yet. For these descendants, we discard the current, unsuccessful main progenitor candidate in favour of the progenitor candidate with the next highest merit, and the first step of the iteration begins anew: All progenitors without a matching main descendant again loop over all their descendant candidates in search for a match. This two-step iteration is repeated until either all descendants have found their match, or have run out of progenitor candidates.

After the iteration finishes, we may have both progenitors as well as descendants which aren't part of a matching pair. We deal with them as follows:

\begin{enumerate}

\item Progenitors that have not found any available descendant will be
  considered to have merged into the descendant candidate with the highest merit. These progenitors are recorded in the merger tree as
  {\it merged progenitors}.  In this case, only one tracer particle is
  kept for future use, the most strongly bound particle in the list of $n_{\rm
    mb}$ tracer particles.  This single particle is referred to as the
  \emph{orphan particle} of the merged progenitor.  It is used to
  check whether the merger event was a final merger or only a
  temporary merger. It is also used to track orphan galaxies.

\item Descendants that have not found any available progenitor will be
  checked against non-consecutive past snapshots. The particles of the
  descendant are compared to the orphan particles in the list of past
  merged progenitors\footnote{There is an option to remove past merged
  progenitors from the list if they have merged into their main
  descendant too many snapshots ago.  By default, however, the
  algorithm will store them all until the very end of the
  simulation.}.  The most strongly bound orphan particle will be used
  to restore the broken link with its main progenitor.  Finally,
  remaining descendants without a progenitor are considered as being
  newly formed.

\end{enumerate}

Finally, we note that the choice to first check all progenitor candidates of descendants and only
merge progenitors into descendants later is how we deal with fragmentation
events (see Section~\ref{sect:frag}) in an attempt to preserve the formation
history of clumps. Effectively, this procedure assigns more weight to a
descendant having progenitor candidates \textit{at all} over the merging of a progenitor into its main descendant candidate, as the merit function would suggest.

    %============================================================================
\section{Testing and Optimizing the Algorithm}\label{chap:tests}
%============================================================================

The current implementation of our merger tree algorithm needs several
free parameters to be chosen by the user. We present in this section
multiple tests that reveal the recommended values for these
parameters.  We use for this a single reference cosmological
simulation and analyze the merger trees we obtained for different set
of parameters.  A similar methodology was used in the Sussing Merger
Trees Comparison Project \citep{SUSSING_COMPARISON,
  SUSSING_CONVERGENCE, SUSSING_HALOFINDER,leeSussingMergerTrees2014a}.
This is why we adopt in this section several tests from this seminal
work, as they are quite efficient at testing the strengths and the
weaknesses of merger tree codes. They also allow for a direct
comparison of our new implementation with many other state-of-the-art
merger tree codes in the community.

%============================================================================
\subsection{Test Suite}\label{chap:testing_methods}
%============================================================================

We now list the different diagnostics we use to characterize the quality of
our merger tree algorithm.

\begin{enumerate}
  \setlength\itemsep{1em}
\item \emph{Length of the Main Branch of a Halo}
  
  The length of the main branch of a halo is simply defined as the
  number of snapshots in which a halo and all its progenitors are
  detected.  A halo at $z=0$ without any progenitors will be
  considered as newly formed and thus will have a main branch of
  length $1$.  If a halo appears to merge temporarily into another and
  re-emerges at a later snapshot, the missing snapshots will be
  counted towards the length of the main branch as if they weren't
  missing. Traditionally, finding long main branches is considered
  as a good thing for a merger tree code.
  
\item \emph{Number of Branches of a Halo}
  
  Another popular quantity is the number of branches of the tree
  leading to the formation of a halo at $z=0$.  The main branch is
  included in this count, thus the minimal number of branches is
  $1$. If a different choice of parameters leads to a reduction of the
  number of branches, it usually corresponds to an increase of the
  average length of the main branches and a smaller number of merger
  events. For example, intuitively if we compare the merger trees where
  we use only one tracer particle per clump to the trees that were built 
  using several hundreds tracer particles per clump, we would expect to 
  be able to detect more fragmentation events with the increased number of 
  tracer particles. If the fragmentation remained undetected, we would 
  instead have found a newly formed clump (the fragment) alongside a 
  merging event. Both the ``newly formed'' fragment as well as its
  progenitor, which is now merged into a descendant, will have shorter
  main branches. Conversely, the descendant will have an increased 
  number of branches compared to the scenario where the fragmentation
  was detected.
  Finally, in the hierarchical picture of structure formation, one
  would expect more massive clumps to have longer main branches and a
  higher number of branches.
  
\item \emph{Logarithmic Mass Growth of a Halo}
  
  The logarithmic mass growth rate of a halo is computed using the following
  finite difference approximation:
  \begin{equation}
    \frac{\de \log M}{\de \log t} \simeq
    \frac{(t_{k+1}+t_{k})(M_{k+1} -M_{k})}{(t_{k+1} - t_k)(M_{k+1} +
      M_{k})} \equiv \alpha_M(k, k+1)
  \end{equation}
  where $k$ and $k+1$ are two consecutive snapshots, with the
  corresponding halo mass $M_k$ and $M_{k+1}$ and times $t_k$ and
  $t_{k+1}$. A convenient approach was proposed by \cite{SUSSING_CONVERGENCE}
  to reduce the range of values to the interval $(-1, 1)$ using the new variable
  \begin{equation}
    \beta_M = \frac{2}{\pi}\arctan(\alpha_M) \label{eq:massgrowth}
  \end{equation}
  Note that we expect the mass of dark matter haloes to increase
  systematically with time.  We also expect in some cases the mass to
  remain constant or even to decrease slightly.  We nevertheless
  expect the distribution of $\beta_M$ to be skewed towards $\beta_M >
  0$.  $\beta_M \rightarrow \pm 1$ imply $\alpha_M \rightarrow \pm
  \infty$, indicating suspiciously extreme cases of mass growth or
  mass loss.
  
\item \emph{Mass Growth Fluctuation of a Halo}
  
  Mass growth fluctuations are defined similarly as
  \begin{equation}
    \xi_M = \frac{\beta_M(k, k+1) - \beta_M(k-1, k)}{2} \label{eq:massfluct}
  \end{equation}
  where $k-1$, $k$, $k+1$ are three consecutive snapshots.  A smooth
  mass accretion history generally leads to $\xi_M \simeq 0$.  Strong
  deviations from zero could indicate an erratic behaviour, indicating
  extreme mass loss followed by extreme mass growth and vice versa.
  Within the standard model of structure formation, this behaviour is
  expected only during major merger events. Otherwise, it might
  indicate either a misidentification by the merger tree code or a
  misdetection by the halo finder.

%\item {\it Displacement statistics}
%  The Sussing Merger Trees Comparison Project also includes a
%  displacement statistic to quantify misidentifications,
%  \begin{equation}
%  \Delta_r = \frac { | \mathbf{r}_{k+1} - \mathbf{r}_k - 0.5
%    (\mathbf{v}_{k+1} + \mathbf{v}_k) (t_{k+1} - t_k) | } {
%    0.5(R_{200,k} + R_{200,{k+1}} + | \mathbf{v}_{k+1} + \mathbf{v}_k
%    | (t_{k+1} - t_k) }
%  \end{equation} 
%  where $\mathbf{r}_{k+1}$, $\mathbf{v}_{k+1}$ and $\mathbf{r}_k$,
%  $\mathbf{v}_k$ are the position and velocity of a clump at snapshot
%  $k+1$ and its progenitor at snapshot $k$, respectively; $t_{k+1}$
%  and $t_k$ are the cosmic times at which the two clumps were defined,
%  and $R_{200}$ is the radius that encloses an overdensity of 200
%  times the critical density $\rho_{c} = \frac{3 H^2}{8 \pi G}$.
%  Values of $\Delta_r > 1$ would indicate a misidentification, so the
%  parameters minimising $\Delta_r$ should be preferred, provided the
%  acceleration is approximately uniform.  However, the obtained
%  $\Delta_r$ for all parameters showed almost no differences and no
%  indication of what parameters should be preferred, which is why the
%  results of the quantification of misidentifications were omitted
%  from this work.
 
\end{enumerate}

Ideally, we should have tested \texttt{ACACIA} on the dataset used in
\citet{SUSSING_COMPARISON} and \citet{SUSSING_HALOFINDER} (S13 and A14
from here on, respectively).  This would
have enabled a direct comparison of the performance to other merger
tree codes.  However, \texttt{ACACIA} was designed to work on the fly
with the \ramses\ code.  Using it as a post-processing tool would
defeat its purpose and as a matter of fact handling other halo 
catalogues  has proven technically impossible. \texttt{ACACIA} is 
tightly coupled to the \phew\ halo finder, and relies heavily on 
already existing internal structures and tools, in particular the 
explicit communications which are necessary for parallelism on 
distributed memory architectures, as well as the structures and 
their hierarchies as they are defined by \phew. Attempting to use 
other halo catalogues would require us to re-write a significant 
portion of the \phew\ halo finder. If we instead used only particle
data, which is possible, we would still find a different halo catalogue
compared to other structure finding codes, and we would still not be
able to do an exact comparison. Furthermore, we also want to demonstrate that the
\phew\ halo finder can be used within the \ramses\ code to produce
reliable merger trees.  For these various reasons, we have decided to
perform the same tests but using our own dataset generated on the fly by
\ramses.

Despite this limitation, we have performed a direct comparison to
other halo finders and merger tree codes using the exact same merger
tree parameters as in A14. The results are
given in Appendix~\ref{app:performance_comparison}.  Our results are
comparable to e.g. the \texttt{MergerTree}, \texttt{TreeMaker} and 
\texttt{VELOCIraptor} tree builders with \texttt{AHF}, \texttt{Subfind},
or \texttt{Rockstar} halo finders as presented in A14, demonstrating that
\texttt{ACACIA} performs similarly than other state-of-the-art tools.

In this section, we would like to explore different parameters and see
how they affect the quality of the merger tree.  Our tests are
performed on a single DMO simulation with $256^3 \approx 1.7\times
10^7$ particles of identical mass $m_p = 1.6\times 10^9\msol$. To enable
a comparison with A14, we adapted the same cosmology
and snapshot output times as them at a comparable, but slightly lower
resolution. The cosmological parameters used are taken from the WMAP-7 
\citep{komatsuSevenYearWilkinsonMicrowave2011}, while the snapshot times 
are identical to the ones used for Millenium Simulation 
\citep{springelSimulationsFormationEvolution2005a}, starting at redshift
50 and being roughly uniformly spaced in log $a$ in 61 steps. 
At redshift zero, the simulation was then continued for 3 further snapshots 
to ensure that the merging events at $z = 0$ are actual mergers and not 
temporary mergers that will re-emerge later.

This choice of spacings between the snapshots was relatively
arbitrarily in the sense that we did not take into account any further
underlying physical considerations that would be important in e.g.
semi-analytical models. For different snapshot spacings, we recommend to 
follow the suggestions found by \citet{SUSSING_CONVERGENCE}:

\begin{itemize}
\item  Sequences of snapshots with very rapidly changing time intervals 
between them should be avoided as they can lead to very poor trees.
\item  Increasing the number of outputs from which the tree is generated 
results in shorter trees. This is because, due to limitations in the input halo 
catalogue, tree-builders may face difficulties caused by the fluctuating center 
and size of the input haloes, and the frequency of detected temporary merging 
events increases with the number of snapshots, resulting in haloes missing from
the catalogue. For merger trees built from
an order of 100 or more snapshots, they recommend using an algorithm capable of
dealing with these problems, which \texttt{ACACIA} is able to do, although at the
moment this patching of missing haloes in the catalogue isn't based on a physical
timescale, but on a user defined number of snapshots.
\item  To facilitate this patching at the end of the simulation, snapshots should 
be generated beyond the desired endpoint. This would entail typically running past 
$z = 0$, as we did with our test suite. 
\end{itemize}

For the clump finder, we have adopted a outer density
threshold of 80$\bar\rho$ and a saddle surface density threshold of
200$\bar\rho$, where $\bar\rho = \Omega_m \frac{3 H^2}{8 \pi G}$ is
the average density.  The minimal mass for clumps is set to 10
particles. Note that for the histogram of the logarithmic mass growth
and mass growth fluctuations, we adopt a threshold for the clump mass
of 200 particles for sake of visibility. Choosing a smaller mass
threshold would indeed give too much weight to small mass, poorly
resolved haloes in our statistical analysis.

%==========================================================================================
\subsection{Varying the Clump Mass Definition}\label{chap:varying_clump_mass_definition}
%==========================================================================================

In our current implementation, there are two important parameters that
can have a strong effect on the halo catalogue (beside the mass and
the density thresholds mentioned earlier) and the corresponding merger
tree.

The first one is the exact definition adopted for the mass of the
sub-haloes in the merit function.  For main haloes, there is no
ambiguity as the mass is defined as the sum of the masses of all
particles contained within the boundary of the halo (set by the outer
density isosurface). This is not the case for sub-haloes, because of the
unbinding process described in Section~\ref{chap:phew}. Indeed,
unbound particles are removed from their original sub-halo and passed
to the parent sub-halo in the hierarchy.  Clump masses are therefore
defined as the sum of the mass of all bound particles.  In the merit
function evaluation, we however consider two different cases to
compute the mass: 1- the mass is equal to the sum of the masses of
only the bound particles, like for sub-haloes or 2- the mass is equal
to the sum of the masses of all particles within their boundaries (set
by the saddle surface with neighbouring clumps), like for main haloes.
In the former case, the mass used in the merit function is identical
to the clump mass. It is referred to as the \exc\ case.  In the latter
case, the mass in the merit function is different that the sub-halo
mass definition but identical to the main halo mass definition.  We
refer to this case as \inc.

The second important definition is the exact boundedness criterion
adopted for the unbinding process.  As discussed in
Section~\ref{chap:phew}, we explored two different cases: When
particles are allowed to leave the outer boundary of their host clump
(and possibly come back later) or when particles are not allowed to
cross the saddle surface during their orbital evolution. In the first
case, we only require the binding energy to be negative, while in the
second case, the binding energy has to be smaller than the
gravitational potential of the nearest saddle point.  We call the
first case \nosad\ and the second case \sad.

\begin{figure*}
  \centering
  \includegraphics[width=\textwidth,keepaspectratio]{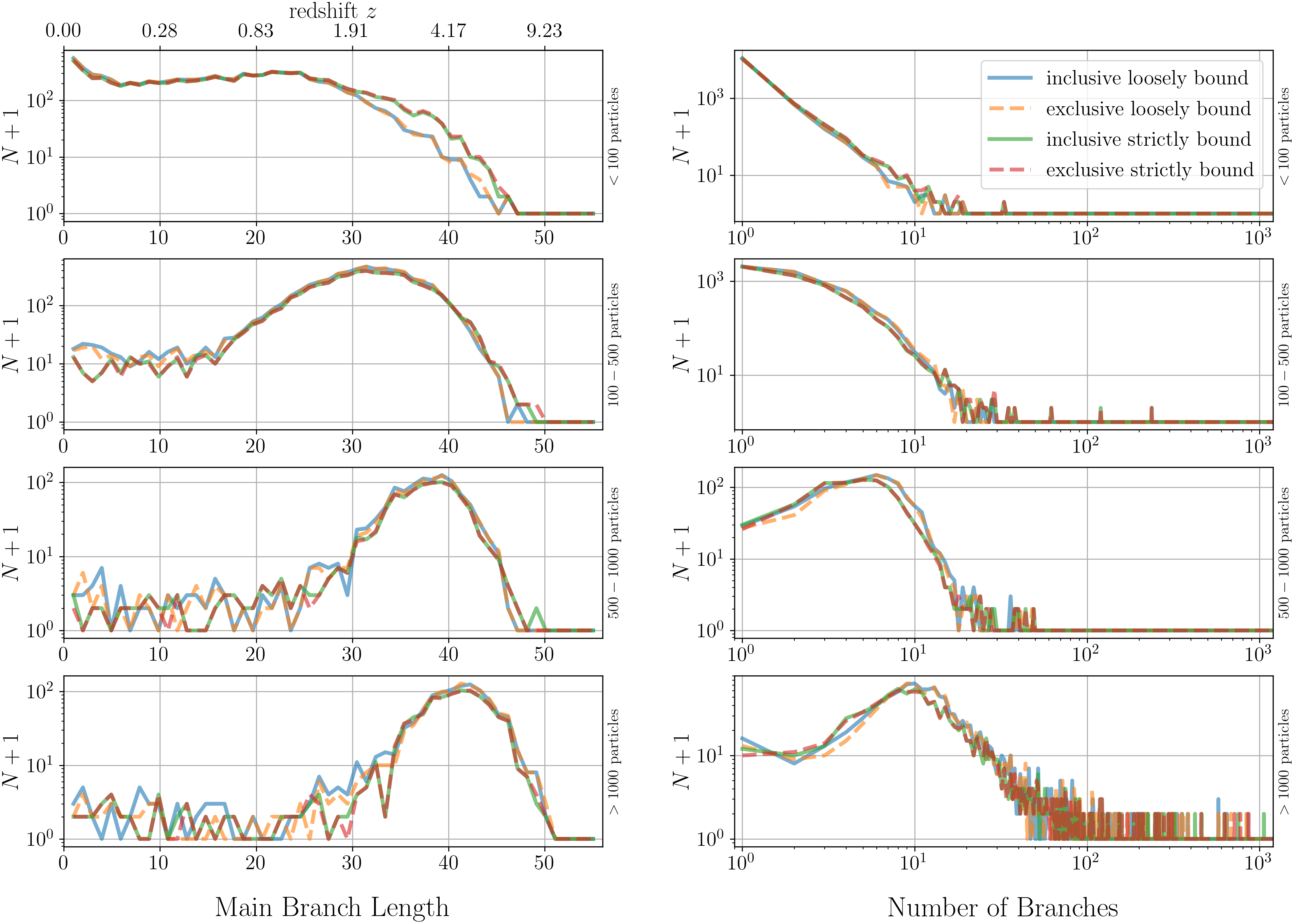}
  \caption{Histogram of the length of main branch (left) and of the
    number of branches (right) for all clumps (halo and sub-halo)
    detected at $z=0$. Each row corresponds to a different range of
    clump masses (expressed in particle numbers): less then 100 (top),
    100-500, 500-1000 and more than 1000 (bottom). We compare these
    histograms for four different cases: whether unbound particles are
    included (\inc) or excluded (\exc) in the evaluation of the merit
    function, and whether bound particles are \nosad\ or \sad.
  }%
  \label{fig:saddle_nosaddle_mbl_nbranch}
\end{figure*}

\begin{figure}
  \centering
  \includegraphics[width=.9\linewidth, keepaspectratio]{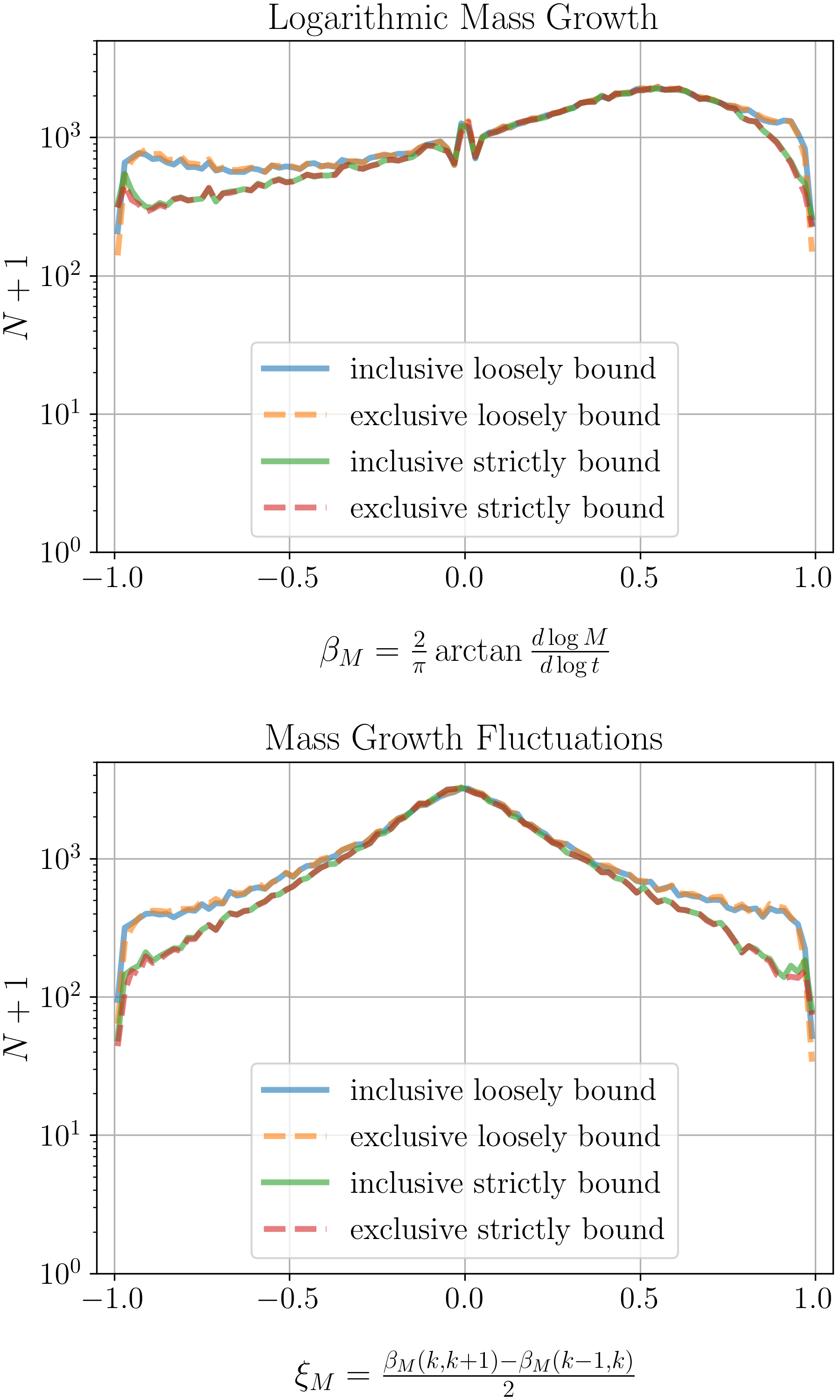}%
  \caption{Histogram of the logarithmic mass growth (top) and the mass
    growth fluctuation (bottom) for all clumps (halo and sub-halo)
    detected in two (top) or three (bottom) consecutive snapshots of
    the simulation and with more than 200 particles. We compare these
    histograms for four different cases: whether unbound particles are
    included (\inc) or excluded (\exc) in the evaluation of the merit
    function, and whether bound particles are \nosad\ or \sad.
  }%
  \label{fig:saddle_nosaddle_masses}
\end{figure}

\begin{table*}
	\caption{
        Average data for all clumps at $z=0$ depending on whether to consider particles which might wander off into another clump as bound (\nosad) or not (\sad).
        The results shown are for the \exc\ mass definition, which show no significant difference to when the \inc\ mass definition is used.
%        The groups I, II, III and IV are defined as clumps that contain less then 100, 100-500, 500-1000 or more than 1000 particles, respectively.
        \label{tab:saddle_nosaddle}
    }
        
%	{\small 
%		\begin{tabular}[c]{l | p{2.8cm} | p{2.8cm} |}
%													&	 \sad\  &   \nosad\ \\ 
%			
%			\hline
%			%
%			total clumps	 						&	 16262	& 	17242 	\\			
%			%		
%%			max number of particles in a clump 		&	 414570	& 	271438 	\\			
%			%		
%			median number of particles in a clump 	&	 83		& 	93 		\\			
%			%
%			\hline
%			%		
%			average main branch length group I		&	23.153	& 	20.527 	\\			
%			%
%			average main branch length group II		&	48.835	& 	48.642 	\\			
%			%		
%			average main branch length group III	&	54.715	& 	54.904 	\\			
%			%
%			average main branch length group IV		&	53.958	& 	56.265 	\\			
%			%
%			\hline
%			%		
%			average number of branches group I		&	1.327	& 	1.230 	\\			
%			%
%			average number of branches group II		&	3.448	& 	3.603 	\\			
%			%		
%			average number of branches group III	&	8.670	& 	8.661 	\\			
%			%
%			average number of branches group IV		&	29.457	& 	28.607 	\\				
%			%
%			\hline	
%		\end{tabular}
%	}

	{\small 
		\begin{tabular}[c]{l | p{2.8cm} | p{2.8cm} |}
													&	 \sad\  &   \nosad\ \\ 
			
			\hline
			total clumps	 													&	 17115	& 	18247 	\\			
			median number of particles in a clump 	&	 77			& 	85 		\\			
			\hline
			average main branch length & & \\
			clumps with < 100 particles			&	14.7	& 	13.0 	\\			
			clumps with 100-500 particles		&	31.4	& 	31.0 	\\			
			clumps with 500-1000 particles	&	37.5	& 	37.3 	\\			
			clumps with > 1000 particles		&	40.7	& 	40.9 	\\			
			\hline
			average number of branches & & \\
			clumps with < 100 particles			&	1.2		& 	1.1 	\\			
			clumps with 100-500 particles		&	2.8		& 	2.8 	\\			
			clumps with 500-1000 particles	&	6.2		& 	6.7 	\\			
			clumps with > 1000 particles		&	25.4	& 	26.1 	\\				
			\hline	
		\end{tabular}
	}
\end{table*}

We now test our algorithm with these four different options for the clump
masses, using the simulation presented in the previous section. Note
that we used here $n_{\rm mb}=200$ tracer particles to identify links
in the merger tree. We will study the impact of this other important
parameter in the next section.

We show in Figure~\ref{fig:saddle_nosaddle_mbl_nbranch} the histogram
of the length of the main branch and the histogram of the number of
branches for each clump (halo and sub-halo) at $z=0$ and for each of
our four different mass definitions. In all four cases, we see that more
massive clumps tend to have longer main branches and a higher number
of branches.  This is also visible from the average length of the main
branch and the average number of branches in different bins of halo
masses given in Table~\ref{tab:saddle_nosaddle}. This is a well-known
property of cosmological simulations in the hierarchical scenario of
structure formation.

Whether clump masses are defined in an \exc\ manner (like for
sub-haloes) or in an \inc\ manner (like for main haloes) in the merit
function has negligible effect on these two statistics. This means
that this a priori large difference in the mass definition of the
merit function has no effect on the linking process of the merger
tree.  The distinction between \sad\ and \nosad\ particles does not
change much for larger clumps but does change the length of the main
branch (and to a lesser extent the number of branches) for small mass
clumps (less than 100 particles). Our first idea was that
\sad\ particles might be better at identifying robust links between
snapshots. It turned out that the main effect of changing the mass
definition from \nosad\ to \sad\ is to reduce the mass of the clump and
to promote them systematically from a larger mass bin to a smaller
mass bin. We see indeed in
Figure~\ref{fig:saddle_nosaddle_mbl_nbranch} and
Table~\ref{tab:saddle_nosaddle} that the number of clumps is reduced
in the large mass bins and increased in the smallest mass bin,
explaining that this change in the mass definition merely transfers clumps
between different bins and affects the statistics accordingly.

A qualitative comparison of the length of the main branches of the most 
massive haloes that we obtain in Figure~\ref{fig:saddle_nosaddle_mbl_nbranch} 
to Figure~3 of A14 shows that our results are in good agreement with what 
the other codes find: The distribution peaks around the length of 45, it 
is about 20 snapshots wide, and there are only few cases with main branch 
lengths below 30. This is in good agreement with what e.g. the 
\texttt{MergerTree}, \texttt{TreeMaker}, and \texttt{VELOCIraptor} tree builders
find in combination with the \texttt{Rockstar} or \texttt{Subfind} halo 
finders. We note that in Figure~3 of A14, both the peak of the
distribution and the maximal value of the main branch lengths they found are
at slightly higher values than ours. We attribute this to the slightly lower 
resolution of our simulations: The first identifiable clumps we find are at 
snapshot 10, leading to a maximal main branch length of 51, compared to 
$\sim 53$ that A14 find. Compared to Figure~3 of S13, the distributions we
find for the lower mass clumps are also in very good agreement. Our high
clump mass distribution however is much narrower around the peak value of 
$\sim 45$. This difference is due to the different halo finders employed,
as is demonstrated in Figure~3 of A14. The \texttt{AHF} halo finder, which
was used in S13, displays the same differences in the distribution of
main branch lengths for nearly all tree codes.
 
We also note in Figure~\ref{fig:saddle_nosaddle_mbl_nbranch} that a few 
large clumps (with mass larger than 500 particles) at $z=0$ have a
main branch length of unity.  These large clumps don't have any
progenitor and thus essentially appeared out of nowhere. As explained
in S13, this effect is present in many state-of-the-art merger tree 
codes and is due to fragmentation events at the periphery of large 
haloes leading to a misidentification of a few rare progenitor-descendant 
links. We have however identified a second culprit, which is the way that
\phew\ establishes substructure hierarchies and the subsequent particle
unbinding. The hierarchy is determined by the density of the density peak 
of each clump: A clump with a lower peak density will be considered lower 
in the hierarchy of substructure. So in situations where two adjacent 
clumps have similarly high density peaks, their order in the hierarchy 
might swap. The unbinding algorithm then strips the particles from the 
sub-haloes that have the lowest level in the hierarchy and passes it on to 
the next level, amplifying the particle loss which these sub-haloes 
experience. This loss of particles is essential here because it prevents 
the algorithm to establish links between progenitors and descendants. 
About half of the main branches that we tracked back in time were cut 
short for this reason: The leaf of the main branch was a sub-halo with 
much fewer particles ($\sim 100$) whose progenitor the algorithm was 
not able to identify and who in subsequent snapshots was found to be 
the main halo, gaining a lot of mass in a very short time. So this issue
arises due to the halo finder, not due to the tree builder.

The histogram of the logarithmic mass growth shown in
Figure~\ref{fig:saddle_nosaddle_masses} is indeed skewed towards
$\beta_M > 0$, demonstrating that clump masses are on average growing.
Based on the shoulder of the histogram around $\beta_M \sim 0.6$ our
results are comparable to those of the \texttt{HBThalo} and \texttt{Subfind}
halo finders in Column~A of Figure~8 of A14 for all tree codes. We do
find more extreme events with $\beta_M \rightarrow \pm 1$, which are
due to the smaller mass haloes and sub-haloes that we used compared
to A14. Indeed, when we apply the appropriate mass thresholds in 
Appendix~\ref{app:performance_comparison}, these extreme events are
significantly reduced. The small wiggle around $\beta_M = 0$ is due to
the discrete particle masses and the linear binning of the histogram.
Adopting a merit function based on the \inc\ or \exc\ mass definition
has here also no effect on the mass growth and mass growth fluctuation
statistics.  At first sight, using the \sad\ instead of the
\nosad\ definition would have led to more robust links and a smoother
mass growth. In Figure~\ref{fig:saddle_nosaddle_masses}, we do see in
the latter case more extreme mass growth around $\beta_M \rightarrow
\pm 1$ and mass growth fluctuations around $\xi_M \rightarrow \pm
1$. We verified that the increase in the number of these extreme
events for the \nosad\ case is in fact due to a larger number of small
sub-haloes that satisfy the adopted mass threshold of 200 particles.
This just means that mass growth statistics is more robust for large,
well resolved haloes, while smaller clumps, closer to the resolution
limit (between 10 and 100 particles; see discussion below) are less
reliable.

In conclusion, whether to use \inc\ or \exc\ mass definitions in the
merit function has no effect on the final merger trees, while using a
\sad\ definition for the clump mass is preferable. We expected the
\sad\ clumps to allow more stable tracking, but the only effect we 
noticed was that it systematically promotes
sub-haloes to lower masses and so naturally selects better resolved, 
higher mass clumps from the halo catalogue.

%===================================================================
\subsection{Varying the Number of Tracer Particles} \label{chap:testing-nmb}
%===================================================================

\begin{table*}
  \caption{Average length of main branch and average number of
    branches for clumps in different mass bins at $z=0$ and for
    varying numbers of clump tracer particles $n_{\rm mb}$.}
  \label{tab:ntracers}
  
  {\small 
    \begin{tabular}[c]{l | p{1cm} | p{1cm} | p{1cm} | p{1cm} | p{1cm} | p{1cm} | p{1cm} |}
      $n_{\rm mb}=$				&	1 		& 	10 		& 	50 		& 	100 	& 200 	& 500 		& 1000 \\
      \hline
      Average main branch length & & & & & & &\\
      clumps with < 100 particles		&	24.2	& 	24.3	& 	23.6	&	23.4 	& 23.2 	& 22.9 	& 22.7  \\	
      clumps with 100-500 particles		&	50.4	& 	50.1	& 	49.5	&	49.1 	& 48.8 	& 48.8 	& 48.8  \\	
      clumps with 500-1000 particles		&	55.2	& 	54.9	& 	53.3	&	54.1 	& 54.7 	& 54.3 	& 54.2  \\	
      clumps with > 1000 particles		&	56.7	& 	54.9 	& 	52.3	&	52.9 	& 54.0 	& 55.8 	& 56.4  \\
      \hline
      Average number of branches & & & & & & &\\
      clumps with < 100 particles		&	1.2	& 	1.3	& 	1.3	&	1.3 	& 1.3  & 1.4 	& 1.4  \\	
      clumps with 100-500 particles		&	2.7	& 	3.0	& 	3.3	&	3.3 	& 3.4  & 3.6 	& 3.6  \\	
      clumps with 500-1000 particles		&	6.6	& 	7.2	& 	8.1	&	8.2 	& 8.7  & 8.9 	& 9.1  \\	
      clumps with > 1000 particles		&	20.4& 	25.2& 	27.3&	28.6 	& 29.5 & 30.4 	& 31.4 \\	
      \hline	
    \end{tabular}
  }
  
  %	\begin{center}
  %		{\small 
  %		\begin{tabular}[c]{l | p{1cm} | p{1cm} | p{1cm} | p{1cm} | p{1cm} | p{1cm} | p{1cm} |}
  %			$n_{mb}=$								&	1 		& 	10 		& 	50 		& 	100 	& 200 	& 500 		& 1000 \\
  %			\hline
  %	%
  %%			total clumps at $z=0$					&	16262	& 	16262	& 	16262	&	16262 	& 16262  & 16262 	& 16262 \\
  %	%		
  %%			max NoP in a clump 		&	414570	& 	414570	& 	414570	&	414570 	& 414570 & 414570 	& 414570  \\ 	
  %	%		
  %%			median NoP in a clump 	&	83		& 	83		& 	83		&	83  	& 83	 & 83 	& 	83  \\	
%	%
%%			\hline
%	%		
%			average MBL group I		&	24.188	& 	24.330	& 	23.567	&	23.353 	& 23.153 & 22.876 	& 22.656  \\	
%	%
%			average MBL group II		&	50.399	& 	50.116	& 	49.472	&	49.122 	& 48.835 & 48.777 	& 48.762  \\	
%	%		
%			average MBL group III	&	55.233	& 	54.863	& 	53.264	&	54.059 	& 54.715 & 54.327 	& 54.165  \\	
%	%
%			average MBL group IV		&	56.690	& 	54.884 	& 	52.345	&	52.900 	& 53.958 & 55.761 	& 56.448  \\
%	%
%			\hline
%	%		
%			average NoB group I		&	1.228	& 	1.305	& 	1.296	&	1.305 	& 1.327  & 1.357 	& 1.367  \\	
%	%
%			average NoB group II		&	2.699	& 	3.062	& 	3.265	&	3.337 	& 3.448  & 3.586 	& 3.596  \\	
%	%		
%			average NoB group III	&	6.625	& 	7.229	& 	8.051	&	8.206 	& 8.670  & 8.914 	& 9.121  \\	
%	%
%			average NoB group IV		&	20.407	& 	25.237	& 	27.288	&	28.554 	& 29.457 & 30.443 	& 31.420  \\	
%	%
%			\hline	
%		\end{tabular}
  %		}

  %	\end{center}	
\end{table*}

\begin{table*}
  
  \caption{Number of dead trees pruned from the merger tree catalogue
    for varying numbers of tracer particles $n_{\rm mb}$ throughout
    all snapshots.  ``LIDIT'' is an abbreviation for ``last
    identifiable descendant in tree''.  For a LIDIT, no descendant
    could have been identified throughout the simulation and
    consequently the corresponding tree is considered dead and pruned
    from the merger tree catalogue. LIDITS are obviously a spurious
    feature of the merger tree algorithm.
    \label{tab:ntracers-pruning}
  }
  
  %	\begin{center}
  {\small 
    \begin{tabular}[c]{l | p{1cm} | p{1cm} | p{1cm} | p{1cm} | p{1cm} | p{1cm} | p{1cm} |}
      $n_{\rm mb}=$													&	1 		& 	10 	& 	50 	& 100 	& 200 	& 500 	& 1000  
      \\
      \hline	
      dead trees pruned from tree catalogue	&	23617	&	15438	&	14467	& 14433 & 14432 & 14432 & 14433 
      \\	
      highest particle number of a LIDIT		&	6418	&	674		&	182		&	182 	& 182 	& 182 	& 182  	
      \\	
      median particle number of a LIDIT			&	19		&	20		&	20		&	20 		& 20 		& 20 		& 20  	
      \\
      LIDITs with >100 particles pruned 		&	493		&	61		&	32		&	28 		& 26 		& 26 		& 26  	
      \\
      \hline
    \end{tabular}
  }
  %	\end{center}	
\end{table*}

\begin{table*}
  \caption{Number of ``jumpers'' (progenitor-descendant links found
    across non-adjacent snapshots) during the entire simulation for
    varying number of tracer particles $n_{\rm mb}$.  }
  \label{tab:jumpers}
  {\small

\begin{tabular}[c]{l | p{1cm} | p{1cm} | p{1cm} | p{1cm} | p{1cm} | p{1cm} | p{1cm} |}
    $n_{\rm mb}=$                   & 1           & 10          & 50          & 100         & 
    200         & 500         & 1000        \\
\hline
    Total Jumpers                   &    14738    &    15500    &    17654    &    18995    &    
    20505    &    20305    &    20407    \\
\hline
    Jumper Progenitors & & & & & & & \\
    clumps with <  100 particles    &    13372    &    14064    &    15696    &    16448    &    
    17233    &    16795    &    16779    \\
    clumps with  100- 500 particles &     1295    &     1353    &     1833    &     2383    &     
    3024    &     3153    &     3214    \\
    clumps with  500-1000 particles &       52    &       60    &       87    &      121    &      
    176    &      251    &      278    \\
    clumps with > 1000 particles    &       19    &       23    &       38    &       43    &       
    72    &      106    &      136    \\
\hline
%\hline
%    Jumper Descendants & & & & & & & \\
%    clumps with <  100 particles    &    13055    &    13849    &    15761    &    16451    &    
%17211    &    16675    &    16646    \\
%    clumps with  100- 500 particles &     1478    &     1512    &     1794    &     2434    &     
%3116    &     3320    &     3344    \\
%    clumps with  500-1000 particles &      136    &       96    &       71    &       82    &      
%131    &      232    &      286    \\
%    clumps with > 1000 particles    &       69    &       43    &       28    &       28    
%&       47    &       78    &      131    \\
\end{tabular}

} %\small
\end{table*}

\begin{figure*}
  \centering
  \includegraphics[width=\textwidth, keepaspectratio]{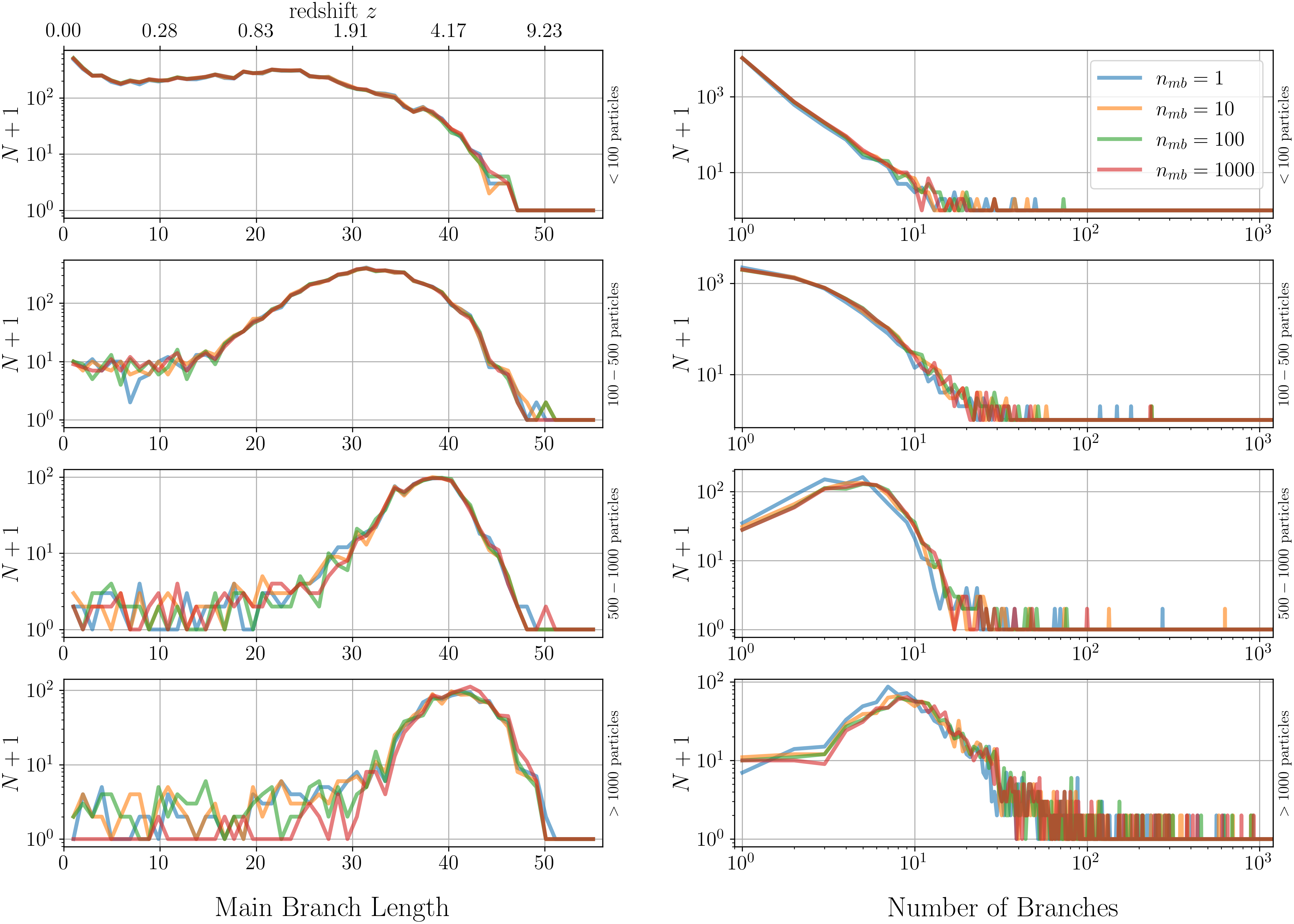}%
  \caption{ Histogram of the length of the main branch (left) and
    histogram of the number of branches (right) for all clumps (haloes
    and sub-haloes) detected at $z=0$ for different numbers of tracer
    particles $n_{\rm mb}$ indicated in the legend.  Each row
    corresponds to a different range of clump masses (expressed in
    particle numbers): less then 100 (top), 100-500, 500-1000 and more
    than 1000 (bottom).  In all cases, we used the \exc\ and
    \sad\ clump mass definitions.
  }%
  \label{fig:ntracers_mbl_nbranch}
\end{figure*}

\begin{figure}
  \centering
  \includegraphics[width=.9\linewidth, keepaspectratio]{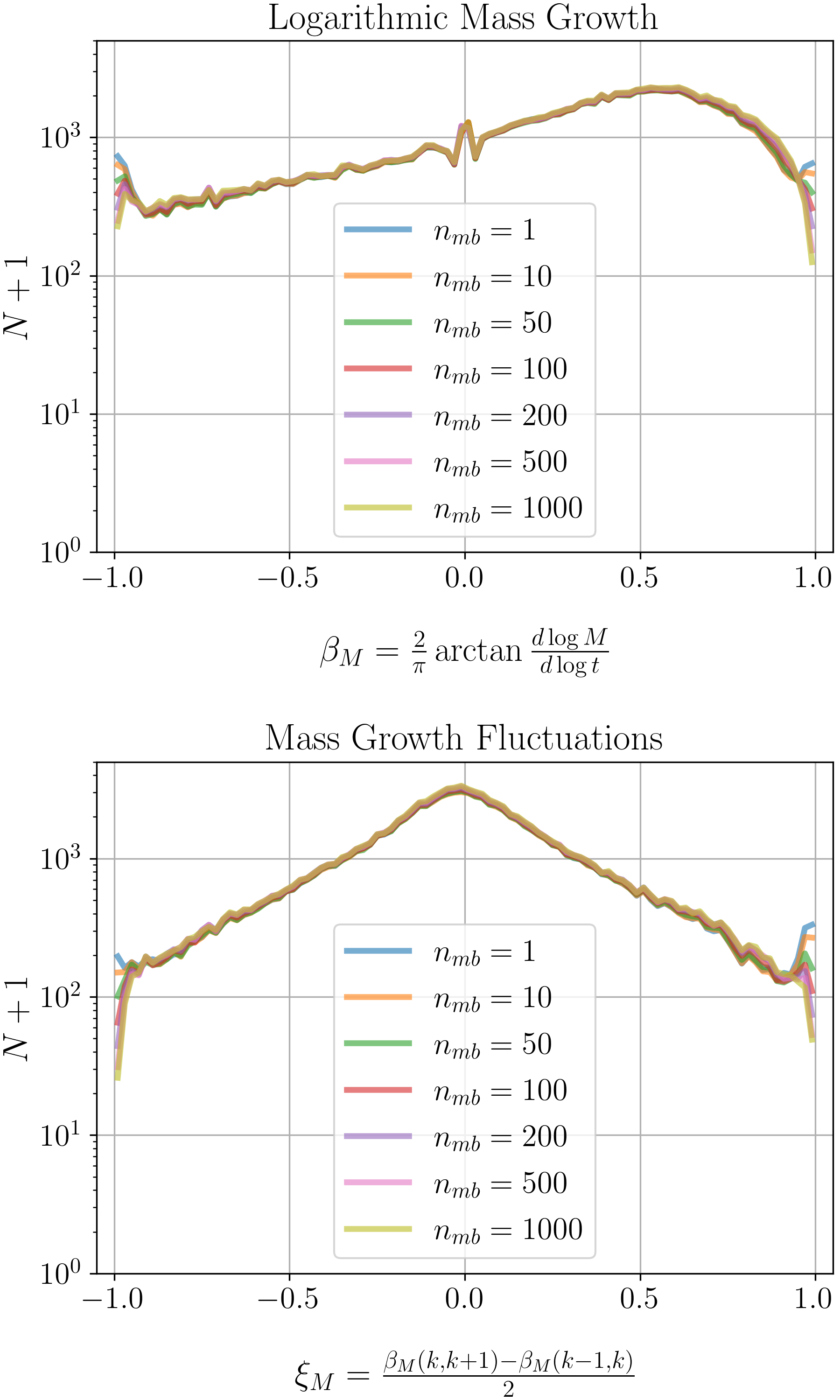}%
  \caption{ Histogram of the logarithmic mass growth (top) and
    histogram of the mass growth fluctuation (bottom) for all clumps
    (halo and sub-halo) detected in two (top) or three (bottom)
    consecutive snapshots of the simulation and with more than 200
    particles.  We compare these histograms for four different numbers of
    tracer particles $n_{\rm mb}$ as indicated in the legend.  In all
    cases, we used the \exc\ and \sad\ clump mass definitions.
  }%
  \label{fig:ntracers_masses}
\end{figure}

\begin{figure}
  \centering
  \includegraphics[width=.9\linewidth, keepaspectratio]{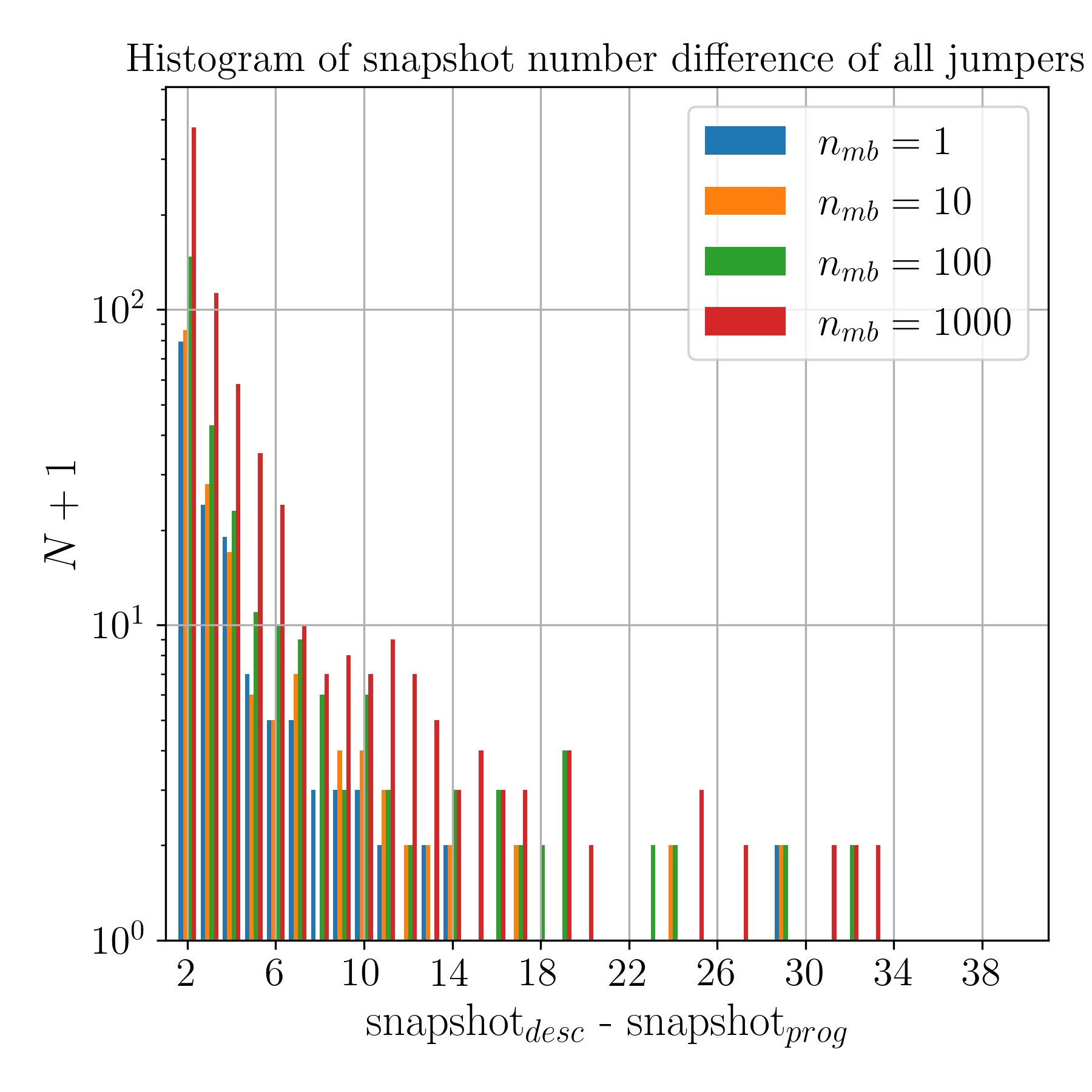}%
  \caption{ Histogram of the difference in snapshot numbers for
    jumpers, i.e.  progenitor-descendant links which are made across
    non-adjacent snapshots.  Only pairs where both the progenitor's
    and the descendant's masses exceed 200 particle masses were
    included in this plot.  We compare these histograms for four
    different numbers of tracer particles $n_{\rm mb}$ as indicated in
    the legend.  In all cases, we used the \exc\ and \sad\ clump mass
    definitions.
  }%
  \label{fig:jumper-distances}
\end{figure}

\begin{figure}
  \centering
  \includegraphics[width=.9\linewidth, keepaspectratio]{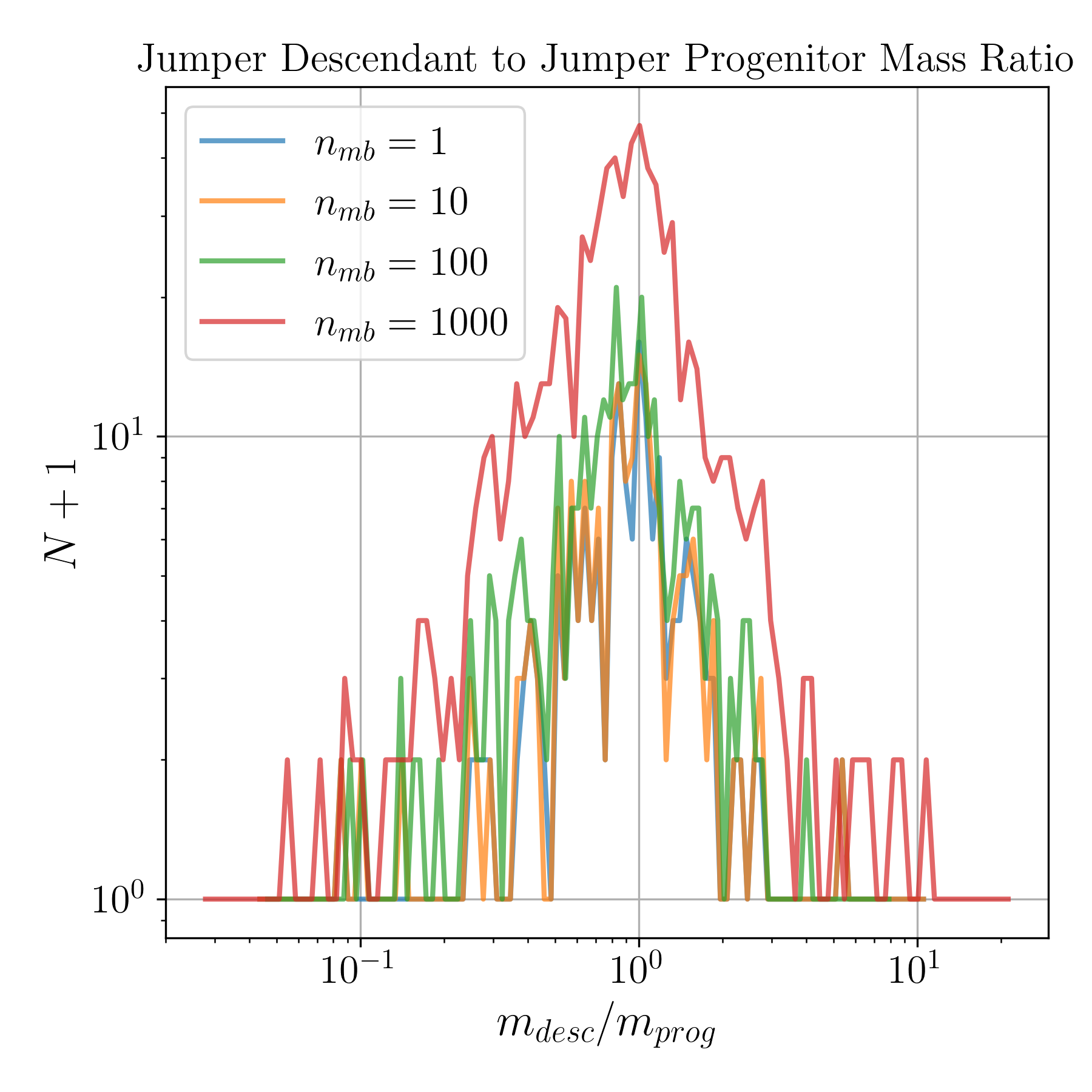}%
  \caption{ Histogram of the ratio of descendant mass to progenitor
    mass for jumpers, i.e.  progenitor-descendant links which are made
    across non-adjacent snapshots.  Only pairs where both the
    progenitor's and the descendant's masses exceed 200 particle
    masses were included in this plot.  We compare these histograms
    for four different numbers of tracer particles $n_{\rm mb}$ as
    indicated in the legend.  In all cases, we used the \exc\ and
    \sad\ clump mass definitions.
  }%
  \label{fig:jumper-mass-ratio}
\end{figure}

In this section, we study the effect of varying the number of tracer
particles on our various diagnostics of tree quality. 
Even though we performed the simulations with
up to 1000 tracer particles, the mass threshold for clumps was always
kept constant at 10 particles. The number of tracer particles per clump 
is an upper limit, not a lower limit. For clumps that contain less than 
$n_{\rm mb}$, this means that they will be traced by every single 
particle they consist of. In effect, we expect that this assigns greater
weight to clumps which are more massive than $n_{\rm mb}$ to be identified 
as the main progenitor, and should hopefully decrease extreme 
mass growths and mass fluctuations. To illustrate, consider for example the 
merging of two clumps with unequal masses, where
all of the tracer particles of both clumps are found inside the resulting
merged descendant. Raising the number of tracer particles above the less
massive clump's particle number in this scenario means that the number of
its tracer particles inside the descendant will remain constant, while
the number of tracer particles stemming from the more massive clump will
increase, and thus raising its merit to be the main progenitor. This is 
also the desired outcome. 
Should however both clumps have masses above $n_{\rm mb}$ particle masses,
then our inclusion of the clump masses in the merit function should 
nevertheless find the more massive clump to be the main progenitor if it
had a mass closer to the resulting descendants mass. So we expect that 
increasing the number of tracer particles should enhance this effect,
and hence lead to at least as smooth mass growths and fluctuations.

We show in
Table~\ref{tab:ntracers} the average number of branches and the
average length of the main branch for all our detected clumps at $z=0$
organised in different mass bins. We see that the effect of the number
of tracer particles used is quite mild. Even with as few as one tracer
particle do we manage to recover the correct average main branch
length. This is also true for small mass haloes, although with a
slightly reduced accuracy. This also validates our orphan particle
technique to track temporary merger events.

On a closer look, the average number of branch is however more
affected by the number of tracer particles. If we look at the two highest
mass bins for clumps in the two bottom rows of Table~\ref{tab:ntracers},
we can see that the average number of branches converges towards
the values of 1000 tracer particles used, and is only slightly lower in
the cases when 200 or 500 tracer particles were used. Comparing these
converged results to the ones with only one tracer particle, 
we loose $\sim 30\%$ of the average number of branches, which are 
links associated to merger events.  This
can be easily explained by the fact that too few tracer particles
cannot be distributed across enough descendant candidates to identify
potential links. It appears that using 100 tracer particles is enough 
to recover most of the otherwise broken links. These
conclusions remain the same after looking at the histogram of the
number of branches and the histogram of the main branch length for the
same clumps in Figure~\ref{fig:ntracers_mbl_nbranch}. Here again, we
see the peak of the histogram of the number of branches being shifted
to the right when increasing the number of tracer particles. We also
see that using 100 tracer particles seems enough to almost recover the
correct distribution.

We now examine the effect of the number of tracers on the mass growth
(and on the mass growth fluctuations) of all our detected clumps
within the entire redshift range. We see in
Figure~\ref{fig:ntracers_masses} that the effect is very weak, except
for the extreme cases $\beta_M \simeq \pm 1$ and $\xi_M \simeq \pm 1$,
corresponding to spurious links in the merger tree. These extreme mass 
growth cases correspond to broken links due to the
small number of tracer particles. Here again, using more than 100
tracer particles seem to get rid of most of these spurious cases.
Note that we include in these histograms only clumps with more than
200 particles.

We now study in details another spurious effect of our merger tree
algorithm, shared by many other merger tree code in the literature,
namely dead tree branches.  A dead branch arises when no descendant
could have been identified after a certain redshift, even after
looking for all subsequent snapshots using the corresponding orphan
particle. Such an event is called a ``Last Identifiable Descendant In
Tree'' or LIDIT.  When such a case occurs, it is customary to prune
the corresponding tree from the tree catalogue. When not enough tracer
particles are used, we expect such spurious dead links to
appear. Table~\ref{tab:ntracers-pruning} shows the statistics of these
LIDITs (or tree pruning events), which confirms that the number of
LIDITs decreases strongly when using more and more tracer
particles. We also show in the same table the typical and maximum mass
of the LIDITs. Interestingly, the maximum mass also strongly decreases
when using more tracer particles. When enough tracer particles are
used, we see that LIDITs are typically less massive than 200
particles. We believe they correspond to poorly resolved clumps that
are subject to all sorts of spurious numerical effects. We found that
taking all LIDITs into account, over 80\% of them were main haloes.
LIDITs containing more than 50 particles however were over 95\% sub-haloes.
This suggests that the number of very low mass LIDITs is dominated
by poorly resolved small clumps in low density environments, since
in overdense environments clumps wouldn't have been identified as main 
haloes, but as sub-haloes instead. Conversely, with increased resolution 
of the clumps, the overwhelming majority of LIDITs are sub-haloes, and as
such in overdense regions. We conclude that a
conservative resolution limit of 200 particles per clump removes all the
LIDITs from our catalogue, as long as one uses more than 200 tracer
particles. 

We finally study a specific aspect of our merger tree algorithm,
namely the possibility to follow the temporary mergers of clumps that
travel through another clumps to emerges later as a distinct object.
We show in Table~\ref{tab:jumpers} the number of these temporary
mergers that we call ``jumpers'' as they represent links across
non-adjacent snapshots that we are able to ``repair'' using orphan
particles. On average, the number of jumpers increases with the
number of tracer particles. This is a similar behavior than for the
number of branches: More tracer particles allows more merger events to
be detected, and every merger event results in a new orphan particle.
More orphan particles allow more non-adjacent descendant candidates
to be found. We here also recommend to use 200 tracer particles as a
compromise between speed and proper detection of jumpers in the
simulation.

We show in Figure~\ref{fig:jumper-distances} the histogram of the
distance in time between the two non-adjacent snapshot of all jumpers
in our merger tree. We see that most jumpers have a distance of only 2
snapshots. They corresponds to clumps traversing another clump and
re-emerging a snapshot later as a distinct halo. We also see in
this histogram that the number of jumpers increases with the number of
tracer particles. But overall, the statistics of the distance between
jumpers is relatively robust, with only very rare cases with a
distance larger than 10 snapshots. Figure~\ref{fig:jumper-mass-ratio}
shows the histogram of the mass ratio between the jumper progenitor
and the jumper descendant.  As expected, it peaks at one, which means
that the mass of the clump that re-emerges in a later snapshot is
close to the mass of the clump that disappeared in an earlier
snapshot. Note that this is not due to the merit function, because for
jumpers we only use a single orphan particle to repair the link. This
is a clear sign that the same clump is identified before and after the
temporary merger.  We also see that the distribution is slightly
skewed toward mass ratio smaller than 1, with values always bounded
between 0.3 and 3. Only a few very rare cases show more extreme mass
ratios. This means that clumps hosting our orphan particles either
preserve their mass (over 2 snapshots) or loose mass (on average),
usually when the time between non-adjacent snapshots increases.

In conclusion, we found that $n_{\rm mb} \simeq 200$ is a safe choice
to obtain robust results for our merger tree algorithm, in light of
the diagnostics we have used in this section.  We also recommend
adopting a conservative mass threshold of 200 particles per clumps to
get rid of a few rare spurious dead branches that would need to be
pruned from the halo catalogue anyway.

    %============================================================================
\section{Application of the Merger Tree Algorithm: Creating a Mock Galaxy Catalogue}
%============================================================================
\label{chap:mock_catalogues}

Now that we  know the optimal parameters to create  a merger tree with
\texttt{ACACIA}, we  use it to  generate a mock galaxy  catalogue.  We
summarize here the main data products generated by our code.
\begin{enumerate}
\item For  every snapshot of the  N body simulation, we  have the full
  clump catalogue  generated by \texttt{PHEW}. Every  clump is uniquely
  classified as a main halo or as a sub-halo.
\item Every dark  matter particle is given the clump  index it belongs
  to, or zero if it belongs to the smooth background.
\item For  each clump, we follow  and store the index  of the $n_{mb}$
  most strongly bound particles, the position of the density peak, the
  clump  bulk  velocities,  centre  of mass,  mass,  and  other  clump
  properties.
\item For each clump, we store the index of the direct progenitors (in
  particular its  main progenitor) and  its peak mass over  its entire
  past formation history.
\item We augment our clump database with orphan particles, storing for
  each of  them the index  of the last  known main progenitor  and its
  peak mass.
\end{enumerate}
To generate  the mock  galaxy catalogue,  we use  the well-established
technique of  Sub-Halo Abundance  Matching (SHAM). This  technique was
introduced  more than  ten  years  ago as  a  surprisingly simple  and
accurate  method  to  populate  a pure  dark  matter  simulation  with
galaxies  with the  correct  clustering statistics 
\citep{valeNonparametricModelLinking2006, shankarNewRelationshipsGalaxy2006, 
conroyModelingLuminositydependentGalaxy2006a}.

Although  several   implementations  of  SHAM  exist   in  the  recent
literature \citep{guoHowGalaxiesPopulate2010, 
wetzelWhatDeterminesSatellite2010, 
Moster2010, trujillo-gomezGalaxiesLCDMHalo2011, 
nuzaClusteringGalaxiesSDSSIII2013,
zentnerGalaxyAssemblyBias2014,
chaves-monteroSubhaloAbundanceMatching2016a} we use here  the variant 
based on the peak clump mass as a proxy for the stellar mass
\citep{reddickConnectionGalaxiesDark2013},          using          the
Stellar-Mass-to-Halo-Mass (SMHM) relation of \cite{Behroozi}.

Using the  peak clump mass  is believed  to mimick the  actual stellar
mass growth of a galaxy, first as a central galaxy when the host clump
was a main halo, then as a satellite galaxy when the halo was accreted
and became  a sub-halo.  After  infall, although the clump  mass might
decrease quickly due to interactions within the parent main halo, this
model assumes  that the  stellar mass in  the galaxy  remains constant
\citep{Nagai}.

Note that in  the SHAM methodology, the merger tree  algorithm plays a
central role.
\begin{enumerate}
\item In order to compute the clump peak mass, we need the entire mass
  growth past history.
\item  In order  to follow  galaxies even  when the  parent clump  has
  dissolved due  to numerical  overmerging, we  need to  follow orphan
  particles and their peak clump mass.
\end{enumerate}

Our parent DMO simulation uses $512^3 \simeq 1.3\times 10^8$ particles
and a box wize of 100 comoving  Mpc with a particle mass resolution of
$m_p \simeq  3.1 \times  10^8\msol$.  The cosmological  parameters are
taken from  the 2015 Planck Collaboration  results \citep{Planck2015},
with  Hubble constant  $H_0  = 67.74$  km s$^{-1}$Mpc$^{-1}$,  density
parameters  $\Omega_m  =  0.309$,  $\Omega_\Lambda  =  0.691$,  scalar
spectral index  $n_s = 0.967$,  and fluctuation amplitude  $\sigma_8 =
0.816$.  The initial conditions  were created using the \texttt{MUSIC}
code \citep{MUSIC}.   As explained  before, the density  threshold for
clump finding was  chosen to be 80 times the  mean background density,
$\bar{\rho} =  \Omega_m \rho_c$, and  the saddle threshold  for haloes
was set to  200$\bar{\rho}$, where $\rho_c = \frac{3  H_0^2}{8 \pi G}$
is the cosmological critical density.

As a  first application of our  merger tree code, we  will now compute
the two point correlation functions and the average radial profiles of
galaxy  clusters in  our simulation.  We will  compare our  results to
observational data, and demonstrate that  this is only when we include
orphan  galaxies  that   our  results  are  in   good  agreement  with
observations.

%============================================================================
\subsection{The Stellar Mass Correlation Function}\label{chap:correlation}
%============================================================================

\begin{figure}
  \centering
  \includegraphics[width=\linewidth, keepaspectratio]{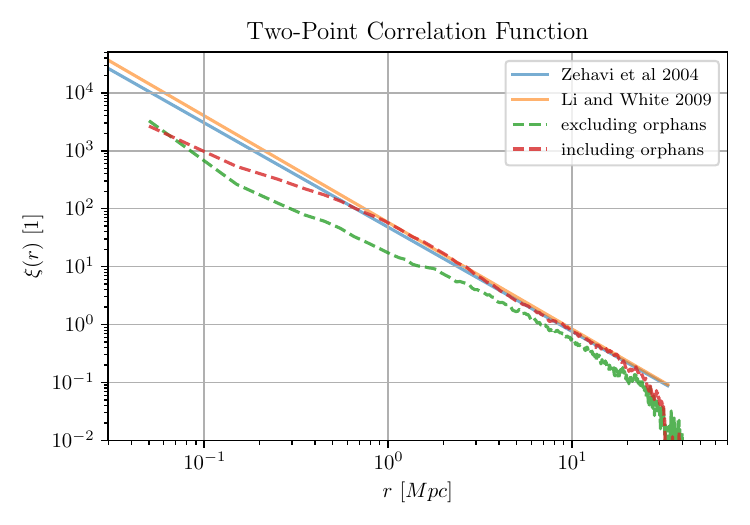}%
  \caption{The  predicted stellar  mass  2-point correlation  function
    (2PCF) $\xi(r)$ of our SHAM  model, including and excluding orphan
    galaxies, compared to  the power law fits of the  observed 2PCF in
    \citet{LiWhite} and \citet{Correlation1}.
  }%
  \label{fig:correlations}
\end{figure}

The  stellar mass  two-point correlation  function (2PCF)  $\xi(r)$ is
computed via  inverse Fourier transform  of the power  spectrum $P(k)$
\citep[e.g.][]{Mo},  which itself  can  be obtained  from the  Fourier
transform    of   the    stellar   mass    density   contrast    field
$\delta(\mathbf{r})$:
\begin{align}
  \delta_\mathbf{k} = \frac{1}{V}\int e^{i\mathbf{kr}} \delta(\mathbf{r}) \de ^3 \mathbf{r}
\end{align}
with
\begin{align}
	\delta(\mathbf{r}) = \frac{\rho(\mathbf{r})}{\langle \rho(\mathbf{r}) \rangle } - 1
\end{align}
Where $\rho(\mathbf{r})$ is the galaxy  stellar mass density field and
$\langle \rho(\mathbf{r}) \rangle$ is  the corresponding mean density,
$V = L^3$ is the volume of our large box on which the density field is
assumed  periodic,  and  $\mathbf{k}=\frac{2\pi}{L}(i_x,  i_y,  i_z)$,
where  $i_x,  i_y,  i_z$  are  integers.   The  Fourier  transform  is
performed using  the FFTW library \citep{FFTW05}.   The power spectrum
$P(k)$ and the 2PCF $\xi(r)$ are given by
\begin{align}
  P(k)    &= V \langle |\delta_\mathbf{k}|^2 \rangle \\
  \xi(r)  &= \frac{1}{(2\pi)^3} \int e^{-i\mathbf{kr}} P(k) \de^3 \mathbf{k} 
\end{align}
The simulation box is divided in  a uniform grid of $1024^3$ cells and
the  stellar mass  is  deposited  on the  grid  using a  cloud-in-cell
interpolation scheme.  The cloud-in-cell scheme consists  of assigning
each  galaxy a  cubic  volume  (``cloud'') the  size  of  a grid  cell
centered on the galaxy's position.  The galaxy stellar mass is assumed
to be uniformly distributed within the  cloud, and is deposited on the
uniform grid cells according to the  volume fraction of the cloud that
resides within each cell.

We only  include galaxies with  masses above $10^9 \msol$.   Using our
adopted SMHM relation,  these galaxies are hosted in  haloes with mass
larger than  $\sim 10^{11} \msol$,  or more than 300  particles.  This
threshold  ensures that  we  only  use well  resolved  clumps for  our
analysis, as discussed in the previous sections.

The predicted 2PCF $\xi(r)$ is shown in Figure \ref{fig:correlations},
and is  compared to  the observational  results of  \cite{LiWhite} and
\cite{Correlation1}.   Note  that  the observed  galaxy  catalogue  is
presented as  complete down  to $10^8 \msol$,  one order  of magnitude
smaller than our  simulated catalogue.  To highlight  the influence of
orphan galaxies,  we have  computed the predicted  2PCF both  with and
without  orphan  galaxies.   Including   orphan  galaxies  produces  a
correlation function in much  better agreement with observations.  Our
theoretical 2PCF obtained  reproduces the observed power  law fit over
two orders  of magnitude  in scale  of $r\sim 0.3  - 25$~Mpc.   On the
smallest  scales, below  0.3~Mpc,  our predictions  are  likely to  be
affected by our limited mass resolution and the grid resolution which 
is used for the Fourier Transform ($\sim 0.1$~Mpc per cell). We have 
verified that using a lower mass threshold for the stellar masses of 
galaxies makes no visible difference in the resulting correlation function.

The  role  of  the  orphan   galaxies  is  particularly  important  on
intermediate scales $\sim  0.2$ Mpc $<\ r \ <  2$ Mpc.  This behaviour
is explained by the fact that  orphan galaxies are located within host
haloes, thus contributing to the  correlations at small distances, the
so-called 1-halo term.  Our conclusion are in agreement  with those of
\cite{crisis}, who  have found that  the inclusion of  orphan galaxies
for mass-based SHAM models improves  the clustering statistics of mock
galaxy catalogues, particularly so at small scales.

%============================================================================
\subsection{Radial Profiles of Satellites in Large Clusters}\label{chap:radial-profiles}
%============================================================================

\begin{figure}
  \centering
  \includegraphics[width=\linewidth, keepaspectratio]{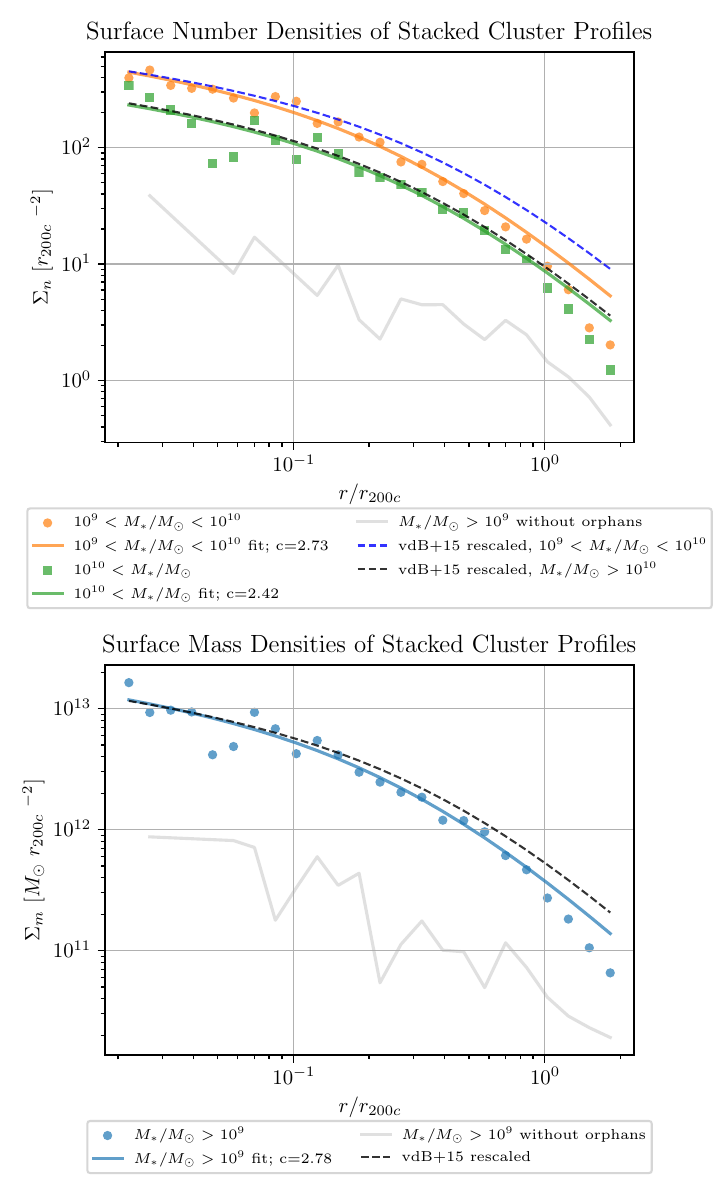}%
  \caption{Upper panel: the light grey  solid line shows the satellite
    number density profile averaged over  the 10 largest haloes in our
    simulation {\it including only  satellite galaxies with a detected
    parent clump with masses above $10^9 \msol$}.   Satellites are  
    then grouped  in two  mass bins
    indicated in  the legend.   The colored  symbols show  the average
    number  density  profile  {\it  including  also  orphan  satellite
      galaxies}.  The colored solid  lines show the corresponding best
    fit projected NFW  profiles, while the  dashed lines  show the best  
    fit projected NFW profiles of the observed sample in
    \citet{vanderburgEvidenceInsideoutGrowth2015}.  Note  that for the
    latter,  we have  renormalized the  surface density  to match  the
    lower median mass of the  simulated sample (see text for details).
    The lower  panel shows the  stellar mass surface  density averaged
    over our  simulated sample. Here  again the light grey  solid line
    shows the  mass density profile  {\it only for  satellite galaxies
      within a detected  clump} while the blue symbols  shows the same
    quantity {\it also including  orphan satellite galaxies}. The blue
    solid  line corresponds  to our  best  fit projected NFW  profile while  the
    dashed line shows the best fit NFW profiles of the observed sample
    in \citet{vanderburgEvidenceInsideoutGrowth2015}.  Here again this
    last profile has  been renormalized to account  for the difference
    in  median mass  between the  simulated and  the observed  cluster
    samples.}
  \label{fig:radial-profiles}
\end{figure}

In this  section, we compute  the radial  profiles of number  and mass
densities  of satellite  galaxies in  the 10  largest clusters  in our
simulation. In  order to  compare to observations,  we will  follow as
closely      as     possible      the     method      described     in
\citet{vanderburgEvidenceInsideoutGrowth2015}, who analyzed 60 massive
clusters between 0.04<$z$<0.26 in  the Multi-Epoch Nearby Cluster Survey
and the Canadian Cluster Comparison Project.

We identified 10  haloes at $z=0$ with the highest  mass.  The mass is
defined here like  in \citet{vanderburgEvidenceInsideoutGrowth2015} as
$M_{200c}$, the  mass included within  radius $R_{200c}$ at  which the
average  enclosed  mass   density  of  the  halo  is   200  times  the
cosmological critical density $\rho_c$.   The $M_{200c}$ masses of our
10 selected  haloes range between  $1.4 \times 10^{14}\msol$  and $4.8
\times 10^{14}\msol$ with a median value of $1.6 \times 10^{14}\msol$.
Due to our limited box size, these  haloes are on the lower end of the
sample observed  in \citet{vanderburgEvidenceInsideoutGrowth2015}, who
reported masses ranging from $0.8 \times 10^{14}\msol$ and $1.6 \times
10^{15}\msol$ with a median value of $8.6 \times 10^{14}\msol$.

We compute  the projected number  density profiles and  projected mass
density  profiles as  follows: For  each  halo, we  first project  the
galaxies  (excluding the  central galaxy)  along each  coordinate axis
obtaining three  images.  We  then compute cylindrical  profiles using
radial shells equally space in log  radius in units of $r_{200c}$.  We
then average  the 30 profiles  (three projections for 10 clusters) to
obtain the final average radial surface density profile.  Each profile
is then fitted to a projected NFW profile \citep{navarroStructureColdDark1996b} using a
standard least square fitting procedure.   We obtain in particular the
concentration parameter that we can compare to the observed value.

The resulting profiles  are shown in Figure~\ref{fig:radial-profiles},
again including  and excluding  orphan galaxies. Orphan  galaxies play
here  also   a  crucial   role,  as   the  profiles   without  orphans
underestimate  the true  value by  an order  of magnitude.   Following
\citet{vanderburgEvidenceInsideoutGrowth2015},   we  adapt   the  same
galaxy  stellar   mass  $M_*$  thresholds   of  $10^9\msol  <   M_*  <
10^{10}\msol$ and $M_* > 10^{10}\msol$  for the surface number density
profile, and  a stellar mass threshold  of $M_* > 10^9  \msol$ for the
surface mass  density profile. It  is worth stressing that  this lower
mass threshold correspond exactly to  the mass resolution limit of our
mock galaxy catalogue.

As already noted above, the median  mass of the simulated and observed
catalogues  widely  differ.  In   order  to  facilitate  a  meaningful
comparison,  we  assume that  the  total  stellar mass  in  satellites
roughly  scales with  $M_{200c}$ in  the halo  mass range  of interest
here. We then adopt the following simple scaling relation
\begin{equation}
\Sigma \propto \frac{M_{200c}}{R_{200c}^2} \propto M_{200c}^{1/3}
\end{equation}
and    rescale    the    observed    average    profile    found    by
\citet{vanderburgEvidenceInsideoutGrowth2015}   using   this   scaling
relation  and  the  ratio  of  the  two  median  masses.  We  plot  in
Figure~\ref{fig:radial-profiles} the corresponding best fit projected NFW
profiles, showing excellent agreement with our simulated results.

Interestingly, the  observed surface density profiles  have a slightly
smaller concentration than  the simulated ones.  Although  we find for
the  number densities  of  satellite galaxies  above $10^{10}\msol$  a
concentration $c  = 2.4$,  in strikingly  good agreement  with $c=2.3$
found  by \citet{vanderburgEvidenceInsideoutGrowth2015}  for the  same
mass range, our results differ for the  mass range $10^9 \msol < M_* <
10^{10}        \msol$:       we        find       $c=2.7$        while
\citet{vanderburgEvidenceInsideoutGrowth2015} found  $c=1.8$. The same
mismatch is  found using the  mass density profile.  we  find $c=2.8$,
while \citet{vanderburgEvidenceInsideoutGrowth2015} found $c=2.0$.

We  believe  that this  mismatch  in  the concentration  parameter  is
consistent  with  the   difference  we  have  in   the  median  sample
mass. Indeed, the  theory predicts a larger  concentration for smaller
mass haloes, roughly in the amplitude observed here \citep{zhaoAccurateUniversalModels2009}. 
We therefore  could in principle improve the agreement
between   our    simulation   results   and   the    observations   of
\citet{vanderburgEvidenceInsideoutGrowth2015}  by  also rescaling  the
radial direction according to the theoretical expectations. We believe
this is  beyond the  scope of this  paper to try  and fit  exactly the
data.

In   addition,    we   believe   there    is   much   more    to   the
story. \citet{vanderburgEvidenceInsideoutGrowth2015} found that larger
mass  satellite galaxies  have  a  significantly larger  concentration
parameter   than    the   low   mass   bin    ($c=2.3$   compared   to
$c=1.8$). Moreover, they have also  found a strong excess of satellite
galaxies compared to the best fit NFW  profile in the centre ($r < 0.1
R_{200c}$).   These observations  are  consistent with  the effect  of
dynamical friction  bringing the more  massive galaxies faster  to the
central regions of  the cluster. Dynamical friction is  expected to be
sufficiently efficient  for the  most massive  sub-haloes, with  a mass
larger than a few  percent of that of the host  halos 
\citep[e.g.][]{binneyGalacticDynamicsSecond2008,Mo}. 
In our case, this  translates into
sub-haloes  more  massive  than  a  few  $10^{12}\msol$  and  satellite
galaxies  more  massive than  a  few  $10^{10}\msol$.  Our  simulation
clearly suffers from  numerical overmerging in this  mass range 
\citep{vandenboschDisruptionDarkMatter2018}, as highlighted  by the 
importance of including  orphan galaxies in
our  methodology.  Moreover,  our  pure DMO  parent simulation  cannot
follow precisely the many baryonic  effects that are needed to predict
accurately the  individual trajectory  of these higher  mass satellite
galaxies.

With these  caveats in mind,  we conclude  that using our  merger tree
code and a  state-of-the-art SHAM method, we can  model reasonably well
the cluster satellite galaxy number density and mass density profiles.

    %=================================================
\section{Conclusion}\label{chap:conclusion}
%=================================================

We presented \texttt{ACACIA}, a new  algorithm to identify dark matter
halo  merger trees,  which is  designed  to work  {\it on-the-fly}  on
systems  with  distributed  memory architectures,  together  with  the
adaptive mesh refinement code \ramses\ with its {\it on-the-fly} clump
finder  \phew.  Clumps  of dark  matter are  tracked across  snapshots
through a user-defined  maximum number of most bound  particles of the
clump  $n_{mb}$.   We found  that  using  $n_{mb} \simeq  200$  tracer
particles is  a safe choice  to obtain  robust results for  our merger
tree  algorithm,  while   not  being  computationally  unrealistically
expensive.  We  also recommend adopting a  conservative mass threshold
of 200  particles per  clump to get  rid of a  few rare  spurious dead
branches that would need to be pruned from the halo catalogue anyway.

Additionally,  we examined  the  influence of  various definitions  of
substructure  properties on  the  resulting merger  trees. Whether  we
define substructures to contain their respective substructures' masses
or not  had negligible effect  on the merger trees.   However defining
particles  to  be  strictly  gravitationally  bound  to  their  parent
substructure  (by requiring  that  particles can't  leave the  spatial
extent of that  substructure) leads to better results,  with much less
extreme  mass growths  and extreme  mass growth  fluctuations of  dark
matter clumps.  We recommend to  use this strictly bound definition as
the  preferred  definition for  robust  merger  trees.  The  resulting
merger  trees  are  in   agreement  with  the  bottom-up  hierarchical
structure formation picture  for dark matter haloes.  The merger trees
of massive  haloes at $z=0$ have  more branches than their  lower mass
counterparts.  Their  formation history  can often  be traced  to very
high redshifts.

Once a progenitor  clump is merged into  a descendant, \texttt{ACACIA}
keeps track of the progenitor's most strongly bound particle, called the
``orphan particle''.  It is possible for a temporarily merged sub-halo
to re-emerge from its host halo  at a later snapshot because it hasn't
actually dissolved  or merged completely,  but only because  it wasn't
detected  by the  clump finder  as a  separate density  peak.  Such  a
situation  is illustrated  in Figure  \ref{fig:jumper-demo}. In  these
cases,  orphan  particles  are  used   to  establish  a  link  between
progenitor  and  descendant   clumps  across  non-adjacent  snapshots.
By  default, \texttt{ACACIA} will track orphans until the end of the
simulation, and orphans are only removed after they have indeed established
a link between a progenitor and descendant and thus have served their
purpose. Nonetheless, the current implementation
offers the option to remove orphan particles after a user defined number
of snapshots has passed. Keeping track of orphan particles indefinitely
might lead to misidentifications of progenitor-descendant pairs and
therefore to wrong formation histories. Our analysis shows however that
matches between progenitor-descendant pairs over an interval greater
than 10 snapshots are quite rare, so we expect this type of
misidentifications to be a negligible issue.

Compared to the test results in \citet{SUSSING_HALOFINDER}, our results
are comparable to e.g. the \texttt{MergerTree}, \texttt{TreeMaker} and
\texttt{VELOCIraptor} tree builders with \texttt{AHF}, \texttt{Subfind},
or \texttt{Rockstar} halo finders as presented in A14, demonstrating that
\texttt{ACACIA} performs similarly to other state-of-the-art tools.
However, the performance is inferior to the one of the \texttt{HBTtree}
algorithm, which together with the \texttt{HBThalo} halo finder follows
structure from one timestep to the next and makes use of this information
when constructing both halo catalogues and trees. Furthermore we have encountered issues, e.g. main branch lengths of massive haloes being cut
short, due to failures in the \phew\ halo finder that we used. In those
cases, substructure changed their order in the hierarchy, and the subsequent
particle unbinding stripped particles from clumps in lower levels in the
hierarchy, preventing \texttt{ACACIA} to establish any links between
progenitor and descendant clumps. The resolution of these issues with the
clump finder will be a high priority in future work, where we plan on
modifying the way \phew\ creates the hierarchies.
For example, we could define the peak hierarchy not based on the peak
density, but rather on the peak mass \citep[e.g. similarly to \texttt{AdapdaHOP},][]{aubertOriginImplicationsDark2004a}. Additionally,
with the structure information from previous snapshots available now through
\texttt{ACACIA}, further improvements can be made by taking this information
into account when constructing the sub-halo hierarchies, in a similar spirit
as \texttt{HBThalo} does.

Orphan particles also serve a second purpose besides tracing disappearing
clumps. If we also
want to produce a mock galaxy catalogue on-the-fly using a dark matter
only simulation, the orphan particles are also used to track orphan
\emph{galaxies}. Those are  galaxies that  don't have  an associated
dark  matter  clump  any  longer  because  of  numerical  overmerging.
If we interpret orphan particles as orphan galaxies, there could be
additional reasons to consider stopping tracking them. For example, the
effects of dynamical friction makes them fall towards the central
galaxies. Once the orphan galaxies have lost enough energy, they may
find themselves in close proximity to the central galaxies, even below
the resolution limit. In these cases, it makes little sense to keep
track of these orphans as individual galaxies. They should rather be
regarded as merged into the central galaxy, and for that reason removed
from the list of tracked orphans. Given that the model we employ
doesn't provide us with the galaxy radii, this approach requires some
form of galaxy-galaxy merging cross-sections to compute the probability
of a collision between galaxies that will result in a galaxy merger.
A different approach that other models use is to estimate the time for
orphan galaxies to merge into the parent structure. This estimate could
be e.g. the dynamical friction time (as is done in \citet{Moster}), or
the fitting formula for the merger timescale of galaxies in cold dark
matter models by \citet{merger_timescales}. These physically motivated
approaches to remove orphans will be the subject of future work, where
we also intend to make use of the tools and methods presented in
\citet{poultonObservingMergerTrees2018} in order to improve our resulting
merger trees and mock galaxy catalogues.

Finally, as a  proof of concept and using the  known formation history
of  dark matter  clumps from  the merger  trees and  a widely  adopted
stellar-mass-to-halo-mass  relation  \citep{Behroozi}, we  generate  a
mock  galaxy  catalogue  from  a dark  matter  only  simulation.   The
influence of the merger trees on  the quality of the galaxy catalogues
is twofold.   First, while the stellar-mass-to-halo-mass  relation can
be directly  applied to  central galaxies  associated to  main haloes,
using the  peak clump  mass for  sub-haloes is  a better  approach for
satellite galaxies.   The reason is  that tidal stripping  of galaxies
inside  a dark  matter halo  sets in  much later  than for  their host
sub-halo \citep{Nagai}.  Second, without properly keeping track of all
merging  events, no  orphan  galaxies  can be  traced,  nor can  their
stellar  mass  be   estimated  through  the  stellar-mass-to-halo-mass
relation  unless  it's  known  from   which  halo  the  orphan  galaxy
originated from, and what properties this halo had in the past.

To highlight  the impact of  the merger trees, we  compute observables
from our  mock galaxy catalogue,  both including and  excluding orphan
galaxies.  Specifically,   we  compute  the  stellar   mass  two-point
correlation  functions   and  radial  profiles  of   projected  number
densities and projected stellar mass densities in galaxy clusters. When orphan
galaxies are included in the analysis, we obtain correlation functions
and radial  profiles in  good agreement with  observations, validating
the different steps in our overall methodology.

%=================================================
\section*{Data Availability}
%=================================================

The data underlying this article will be shared on reasonable request
to the corresponding author. The \ramses\  code is  publicly available
and  can be  downloaded from \url{https://bitbucket.org/rteyssie/ramses/}.
Instructions  on how to use \texttt{ACACIA} and \texttt{PHEW} during a
simulation can be found under
\url{https://bitbucket.org/rteyssie/ramses/wiki/Content}.

%=================================================
\section*{Acknowledgements}
%=================================================

MI would like  to thank B. Roukema for  helpful suggestions concerning
the history  of the application  of merger trees in  the astrophysical
context, S.   Avila for discussions  on details of the  Sussing Merger
Tree Comparison  project, the referee C. Power for his suggestions
which improved this paper, and Y. Revaz for  his support in many  ways.
This  work  was supported  by the  Swiss
National Supercomputing  Center (CSCS)  under projects s1006  and uzh5
and  by the  Swiss  National Science  Foundation  (SNF) under  project
72535 ``Multi-scale multi-physics models of galaxy formation''.  This
work        made        use        of        the        \textsc{numpy}
\citep{harrisArrayProgrammingNumPy2020}       and       \textsc{scipy}
\citep{virtanenSciPyFundamentalAlgorithms2020}  python  libraries  for
the      data      analysis      and      the      \textsc{matplotlib}
\citep{hunterMatplotlib2DGraphics2007}  python  library  for  plotting
tools.

%%%%%%%%%%%%%%%%%%%%%%%%%%%%%%%%%%%%%%%%%%%%%%%%%%


    
    \bibliographystyle{mnras}
    \bibliography{references} % if your bibtex file is called references.bib
    
    \appendix
   	%========================================================================================================
\section{Tree Statistics Using \citet{SUSSING_HALOFINDER} Selection Criteria}\label{app:performance_comparison}
%========================================================================================================

\begin{table}
\centering
\caption{
	Comparison of simulation and evaluation parameters used in this work and of A14, where the parameters of the latter have been converted using $h = 0.704$.
	$m_m$ is the mass threshold for main haloes, $m_s$ is the mass threshold for sub-haloes.
	\label{tab:parameter-comparison}
}
	\begin{tabular}[c]{l l l}
																	&	This work		&	A14 \\
		\hline
		particle mass	[$10^9 \msol$]	&	$1.55$			& $1.32$						\\
		particles used								& $256^3$			& $270^3$ 					\\
		box size [Mpc/h]							& $62.5$			& $62.5$						\\
		snapshots until $z = 0$				& 62					& 62								\\				
		$m_m$ [$10^{12} \msol$]				&	$1.35$			& $1.12$ - $1.37$		\\
		$m_s$ [$10^{11} \msol$]				&	$4.03$			& $4.26$ - $9.72$		\\
		\hline
	\end{tabular}
\end{table}

\begin{figure}
	\centering
	\includegraphics[width=.95\linewidth, 
	keepaspectratio]{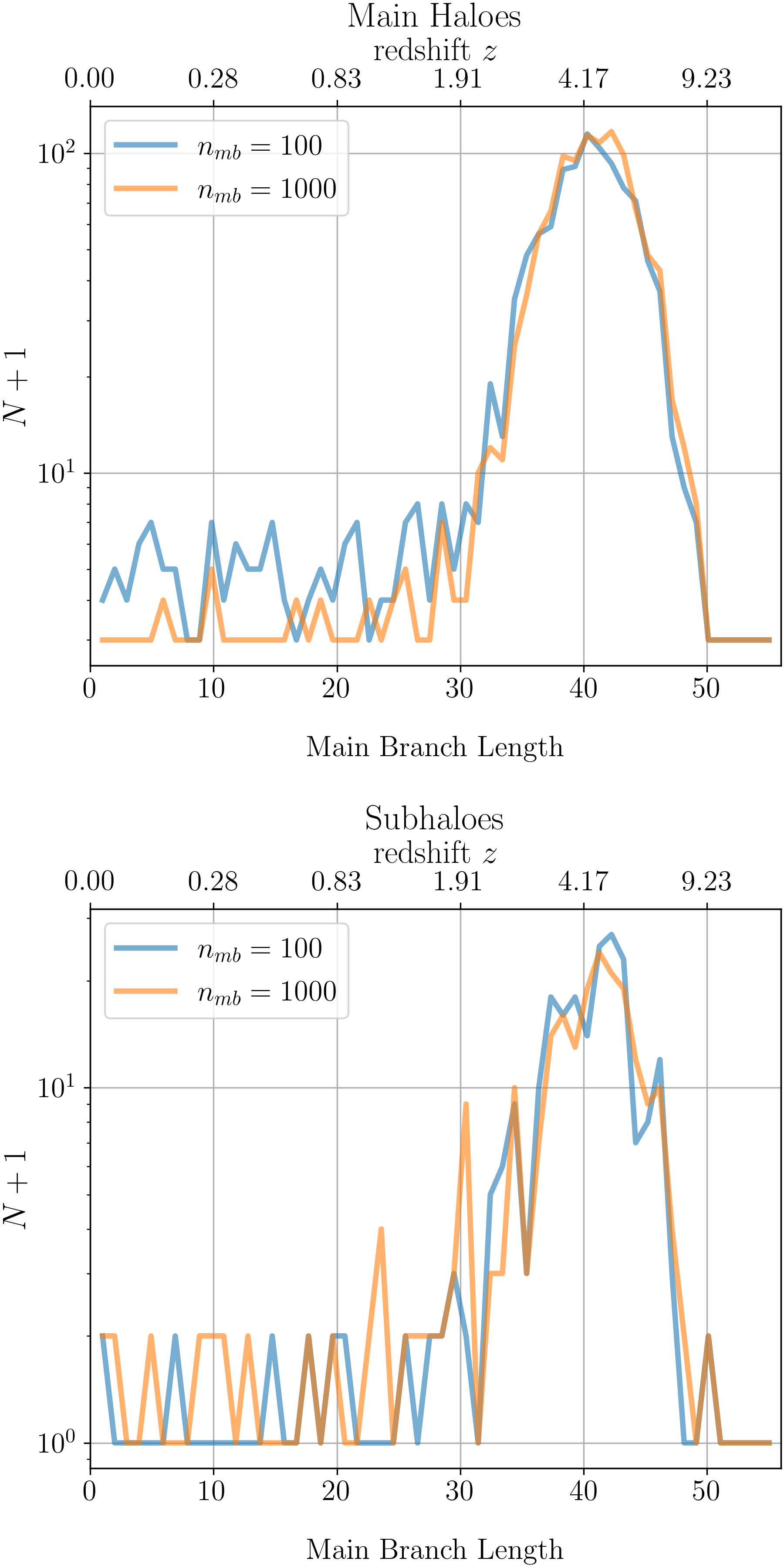}%
	\caption{
		Histograms of the length of the main branch. 
		The top plot shows the length of the 1000 most massive haloes at $z = 0$, the bottom plot shows 
		the length of the 200 most massive sub-haloes for $n_{mb} = 100$ and $1000$ tracer particles 
		per clump.
	}%
	\label{fig:sussing-branch-lengths}
\end{figure}

\begin{figure}
	\centering
	\includegraphics[width=.9\linewidth, keepaspectratio]{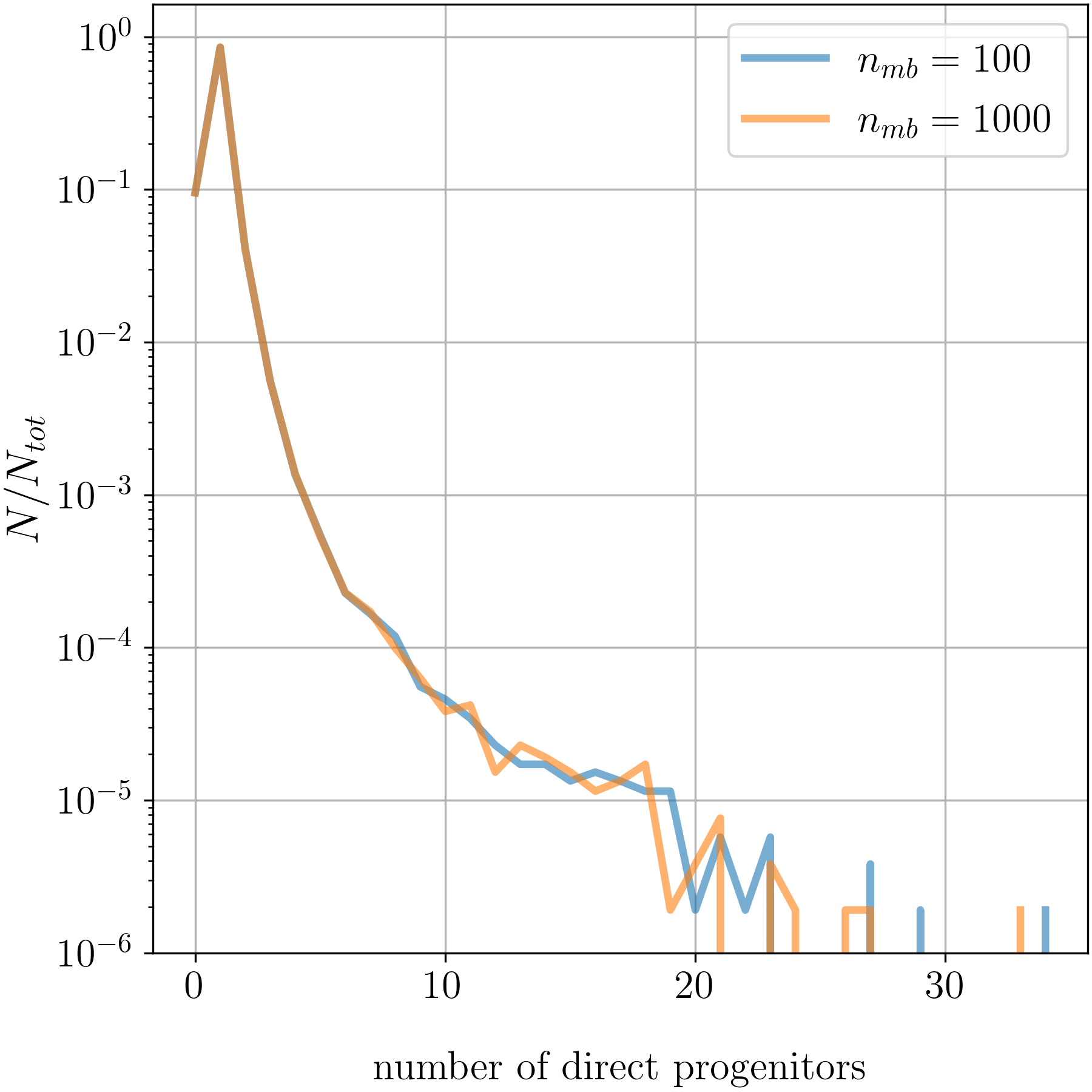}\\%
	\caption{
		Histogram of the number of direct progenitors for all clumps from $z = 0$ to $z = 2$ for $n_{mb} = 100$ and $1000$ tracer particles per clump.
		The histogram is normalized by the total number of events found.
	}%
	\label{fig:sussing-branching-ratio}
\end{figure}

\begin{figure*}
	\centering
	\includegraphics[width=\textwidth, keepaspectratio]{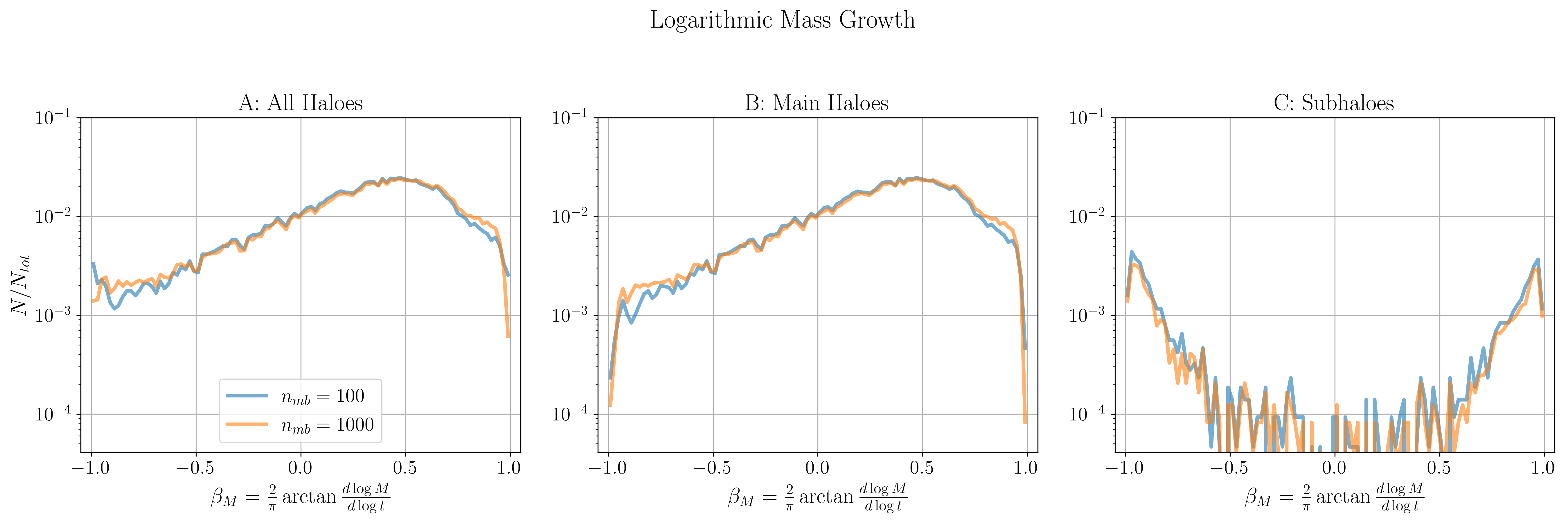}%
	\caption{
		Logarithmic mass growth for haloes and sub-haloes satisfying the mass thresholds.
		Group $A$ contains clumps that are either haloes or sub-haloes in consecutive snapshots $k$ and $k+1$ with masses $m \geq m_{m}$.
		Group $B$ contains clumps that are only haloes in two consecutive snapshots with mass above $m_{m}$, group $C$ contains only clumps that were sub-haloes in two consecutive snapshots with mass greater than $m_{s}$.
		The histogram is normalized by the total number of events found for group $A$.
	}%
	\label{fig:sussing-mass-growth}
\end{figure*}

\begin{figure*}
	\centering
	\includegraphics[width=\textwidth, keepaspectratio]{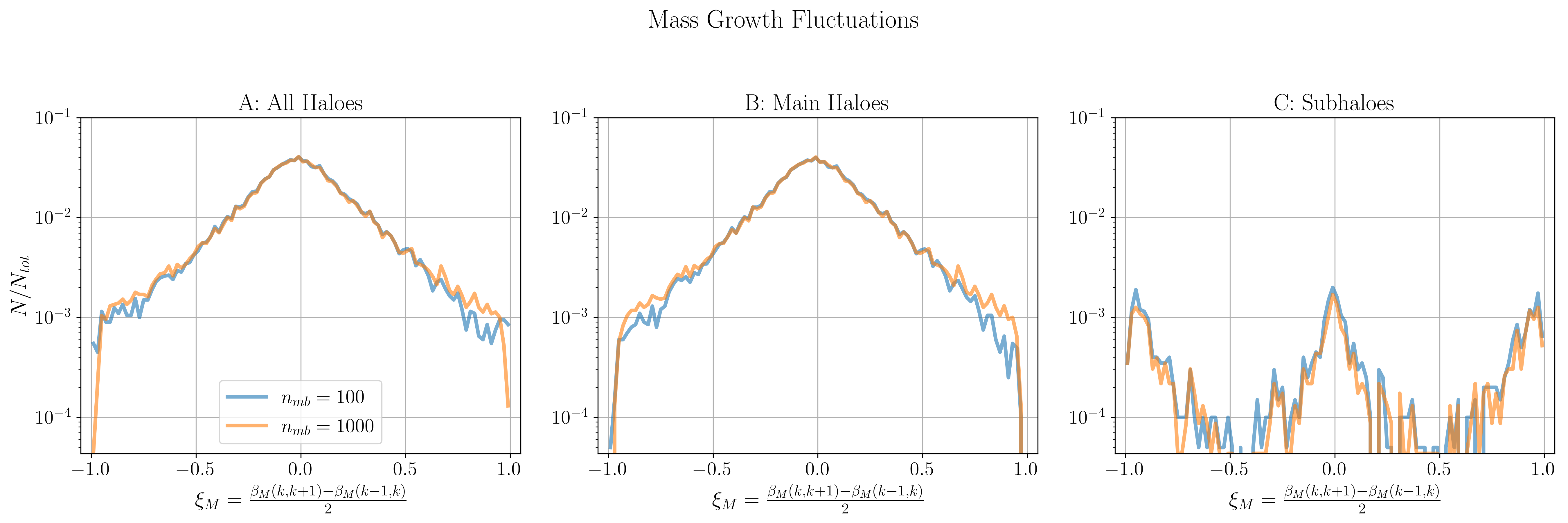}\\%
	\caption{
		Histogram of mass growth fluctuations for haloes and sub-haloes satisfying the mass thresholds.
		Group $A$ contains clumps that are either haloes or sub-haloes in three consecutive snapshots with masses $m \geq m_{m}$.
		Group $B$ contains clumps that are only haloes in three consecutive snapshots with mass above $m_{m}$, group $C$ contains only clumps that were sub-haloes in three consecutive snapshots with mass greater than $m_{s}$.
		The histogram is normalized by the total number of events found for group $A$.
	}%
	\label{fig:sussing-mass-fluct}
\end{figure*}

In this section, the merger tree statistics introduced in Section \ref{chap:tests} when following the selection criteria that are used in \citet{SUSSING_HALOFINDER} (A14 from here on) are presented.
Ideally, \texttt{ACACIA} should be tested on the same datasets and halo catalogues used in the Comparison Project to enable a direct comparison to the performance of other merger tree codes. 
However, since \texttt{ACACIA} was designed to work on the fly, using it as a post-processing utility would defeat its purpose.
Furthermore, \texttt{ACACIA} is not necessarily compatible with other definitions of haloes and sub-haloes.
But most importantly, we also want to demonstrate that the halo finder \phew\ can be used to produce reliable merger trees. 
So instead, the tests are performed on our own datasets and halo catalogues, which are described in section \ref{chap:testing_methods}.
A comparison of the used parameters of our simulations and the ones used in A14 is given in Table \ref{tab:parameter-comparison}.
In the following, the results for $n_{mb} = 100$ and $n_{mb} = 1000$ are shown.
Like before, when the influence of the number of tracer particles was investigated, the \sad\ parameter and the \exc\ mass definition were used.

The difference to the results presented in Section \ref{chap:tests} is that the mass thresholds are set such that only the 1000 most massive main haloes and only the 200 most massive sub-haloes at $z = 0$ are included.
This gives effective mass threshold $m_{m} = 1.35 \times 10^{12} \msol$ and $m_{s} = 4.0 \times 
10^{11}\msol$, which are on one hand comparable to the mass thresholds applied in A14 (Table 
\ref{tab:parameter-comparison}), but already show differences in the resulting halo catalogue.
\phew\ finds a mass threshold for main haloes that is close to the upper limit found in A13, 
but a lower mass threshold for sub-haloes.
This is consistent with the fact that the \sad\ parameter was used:
Unbound particles are passed on to substructure that is higher up in the hierarchy, and the unbinding is repeated until the top level, which are the main haloes, is reached.
The more strict unbinding criterion tends to assign more particles to the main haloes and remove them from sub-haloes, which is reflected in the mass thresholds.
Indeed, using the \nosad\ parameter instead leads to $m_m = 1.27 \times 10^{12}\msol$ and $m_s = 
1.92 \times 10^{12}\msol$.

The length of the main branches for haloes and sub-haloes individually are shown in Figure 
\ref{fig:sussing-branch-lengths}. Compared to Figure 3 of A14, we note the following similarities 
and differences:
\begin{enumerate}
	\item \texttt{ACACIA} finds some main haloes with short (< 10) main branches.
		In A14, his only happens for the \texttt{JMerge} and \texttt{TreeMaker} tree builders
		regardless	of the halo finder employed, and the \texttt{MergerTree}, \texttt{SubLink}, 
		and \texttt{VELOCIraptor} tree makers for \texttt{AHF} and \texttt{Subfind} halo finders.
	\item Like in nearly all cases in A14, the distribution of main branch lengths for
		main haloes peaks at high numbers and the bulk of the distribution is about 20 snapshots
		wide. The peak of the main branch length distribution of \texttt{ACACIA} is at 40,
		while in most cases in A14, it's around 45 with the exception of the \texttt{AHF}
		halo finder. This indicates that \texttt{ACACIA} and \phew\ result in on average 
		somewhat smaller main branch lengths than other codes. However, we also find the 
		maximal main branch length of 50, while nearly all combinations of
		halo finders and tree makers in A14 find higher values. Both these differences can be 
		explained by the slightly lower resolution that we used in our simulations, where we 
		found no identifiable clumps before snapshot 11, which corresponds to a maximal main 
		branch length of 51. 
	\item The main branch lengths for sub-haloes also follow the trends noted above. Additionally,
		the distribution is flat for main branch lengths smaller than 30, which is not the case
		for \texttt{Rockstar} and \texttt{AHF} sub-haloes in A14 for nearly all tree builders. Instead,
		their main branch lengths peak again around unity.
\end{enumerate}
In summary, regarding the main branch lengths of main haloes, \phew\ and \texttt{ACACIA} perform
comparably to \texttt{Subfind} and \texttt{Rockstar} halo finders and \texttt{MergerTree}, 
\texttt{TreeMaker}, and \texttt{VELOCIraptor} tree builders. Concerning sub-haloes, the results are
most closely to \texttt{Subfind} sub-haloes and the same tree builders as for the main haloes. This 
is not surprising, because \texttt{Subfind} employs a similar definition of substructure being 
arbitrarily shaped self-bound structure that is truncated at the isodensity contour that is 
defined by the density saddle point between the sub-halo and the main halo. 

In Figure \ref{fig:sussing-branching-ratio} the number of direct progenitors for all clumps between 
$z = 0$ and $z = 2$ are shown. Comparing to Figure 5 of A14, \texttt{ACACIA} gives very comparable 
results: $\sim 10^{-1}$ haloes have no direct progenitor, almost all have one, and the distribution 
follows an exponential decay with the maximal number of direct progenitors lying around 20-25, save 
for a very few outliers. Many tree makers and halo finders in A14 exhibit the same kind of 
behaviour, particularly so for the \texttt{AHF}, \texttt{Subfind}, and \texttt{Rockstar all} halo 
finders in Figure 5 of A14.

For the logarithmic mass growth (Figure \ref{fig:sussing-mass-growth}) and the mass growth fluctuations (Figure \ref{fig:sussing-mass-fluct}), the statistics are separated into three groups.
Group $A$ contains clumps that are either haloes or sub-haloes in \emph{consecutive} snapshots $k$ 
and $k+1$ with masses greater than the mass threshold $m_m$ in both snapshots. Group $B$ contains 
clumps that are exclusively main haloes in two consecutive snapshots with mass above $m_{m}$, group 
$C$ contains only clumps that were sub-haloes in two consecutive snapshots with mass greater than 
$m_{s}$. We follow clumps of the $z = 0$ snapshot along the main branch only.

The logarithmic mass growth resulting from \texttt{ACACIA} follows the general trend that the tree 
makers in A14 exhibit too. The growth for groups $A$ and $B$ increases steadily and peaks around 
$\beta_M \sim 0.5$, where the peak is $\sim2 \times 10^{-2}$. For $n_{mb} = 100$, the extreme mass 
loss  with $\beta_M = -1$ increases for group $A$, which is an undesirable property, but is also 
exhibited by \texttt{Sublink} in A14. For $n_{mb} = 1000$, it drops to about $10^{-3}$ ($10^{-4}$ 
for group $B$), which is comparable behaviour to \texttt{MergerTree}, \texttt{Sublink}, and 
\texttt{VELOCIraptor}, particularly so in combination with \texttt{AHF} and \texttt{Subfind} halo 
finders. Group $C$, containing only mass growths of clumps that have been sub-haloes in two 
consecutive snapshots, shows a distribution peaking around extreme mass growths $\beta_M 
\rightarrow \pm 1$ at $\sim 5 \times 10^{-3}$, which can again be seen in A14 for almost all tree 
makers, albeit not for all halo finders. More noticeably, almost no sub-haloes are found with $-0.5 
< \beta_M < 0.5$ with \phew\ and \texttt{ACACIA} in Figure \ref{fig:sussing-mass-growth}.
This is due to the fact that once a halo is merged into another, it quickly loses its 
outer mass due to the strict unbinding method used here. However, the distribution that 
\texttt{ACACIA} finds displays some differences with respect to the results in A14. Firstly,
we find almost no mass growth with $-0.5 < \beta_M < 0.5$, similar only to the results of 
\texttt{JMerge} in A14. This is partially due to the strict selection criteria used for this 
analysis: We only include clumps that are classified as sub-haloes in two consecutive snapshots and 
satisfy the mass threshold $m_s$ in both snapshots as well. In particular, this excludes all 
non-adjacent ``jumper'' links that we find, since we don't modify the halo catalogue like e.g. 
\texttt{ConsistentTrees}. Furthermore, we employ the \sad\ unbinding criterium, which strips more
particles from sub-haloes and assigns them to main haloes, leaving the halo catalogue with fewer 
sub-haloes that satisfy the mass threshold. When we instead use the \nosad\ 
criterium, we find that the distribution is on average around $3 \times 10^{-4}$ for $-0.5 < 
\beta_M < 0.5$, albeit noisy, which is in good agreement with most halo finders and tree builders 
in A14. 
Secondly, the distribution \texttt{ACACIA} finds looks remarkably symmetric w.r.t. $\beta_M = 0$.
While e.g. \texttt{MergerTree} and \texttt{TreeMaker} trees with \texttt{AHF}, \texttt{Subfind}, 
and \texttt{Rockstar} haloes find peaks close to $\beta \pm 1$ of similar height, they are also 
always have distributions skewed towards mass losses $\beta_M < 0$ and the peaks at $\beta_M = -1$ 
higher than the one at $\beta_M = 1$. 

We found that the reason why our distribution looks so 
symmetrical is due to the particle unbinding method and the way sub-halo hierarchies are 
established in \phew, similarly to what we have found to be a reason for the short main branch
lengths in Section~\ref{chap:varying_clump_mass_definition}. The hierarchy is determined by the 
density of the density peak of each clump: 
A clump with a lower peak density will be considered lower in the hierarchy of substructure. So in 
situations where two adjacent sub-haloes have similarly high density peaks, their order in the 
hierarchy might change in between two snapshots due to small changes. The unbinding algorithm then 
strips the particles from the sub-haloes that have the lowest level in the hierarchy and passes it 
on to the next level, amplifying the mass loss which these sub-haloes experience. If in the next 
snapshot the order in the hierarchy for these two clumps are inverted, the clump which experienced 
a mass loss previously will now experience a strong mass growth and vice versa. In 
Figure~\ref{fig:sussing-mass-growth} such an oscillation over two snapshots will simultaneously add 
a strong mass growth and a strong mass loss twice in place of a net smoother mass loss, leading to 
the symmetry of the distribution. We verified that about 10$\%$ of strong mass growth events with 
$\beta_M > 0.75$ are also accompanied by the respective sub-haloes increasing their level in the 
hierarchy. Similarly, about 10$\%$ of strong mass loss events with $\beta_M < -0.75$ are 
accompanied by the respective sub-haloes decreasing their level in the hierarchy.

The mass growth fluctuations (Figure \ref{fig:sussing-mass-fluct}) of \texttt{ACACIA} share the 
general trend with the ones from Figure 8 in A14, in that they peak around $\xi_M = 0$ and decrease 
outwards towards $\xi_M = \pm 1$. In A14, in all cases groups $A$ and $B$ peak just below 
$10^{-1}$, while our results peak around $4 \times 10^{-2}$. However, similarly to the results of 
e.g. \texttt{Sublink}, \texttt{TreeMaker}, and \texttt{VELOCIraptor} with the \texttt{AHF} or 
\texttt{Subfind} halo finders, the distribution around $\xi_M \sim \pm 0.5$ drops to $\sim 5 \times 
10^{-3}$, and then continues dropping below $10^{-4} - 10^{-3}$ at $\xi_M \sim \pm 1$. 
%For $n_{mb} = 100$, the group $A$ distribution rises again for $\xi_M \sim 1$, as it does for 
%\texttt{TreeMaker} and \texttt{Sublink} tree makers with \texttt{AHF}. 
Group $B$ shows a steeper drop around the extreme values $\xi_M \sim \pm 1$ compared to group $A$, 
dropping below $10^{-4}$ at these values, similarly to the behaviour of many tree makers and halo 
finders in A14.
The sub-halo group $C$ of this work shows three main peaks, around $-1$, $0$, and $1$.
These peaks also appear in the A14 results.
However, the peaks at the extreme values in A14 are lower than the ones of this work, while the 
peaks around $0$ is higher. The reason why these peaks are so pronounced in our results is the same 
as for why the mass growths in Figure~\ref{fig:sussing-mass-growth} is remarkably symmetric 
compared to others: it's sub-haloes and their respective sub-sub-haloes switching their order in the 
substructure hierarchy repeatedly and the particle unbinding algorithm stripping particles from the 
lower level substructure and assigning it to the higher level substructure.
The missing values around $\xi_M \sim \pm 0.5$ that were also seen in the mass growth in Figure 
\ref{fig:sussing-mass-growth} remain unsurprisingly, and are mitigated if the \nosad\ unbinding 
criterion is applied instead.
Similar distributions are obtained by e.g. the \texttt{AHF} and \texttt{Rockstar} halo finders in 
combination with the \texttt{MergerTree}, \texttt{JMerge}, \texttt{Sublink}, and \texttt{TreeMaker} 
tree builders.

In summary, we find that \texttt{ACACIA} and \texttt{PHEW} produce merger tree statistics which are 
similar to what multiple other state-of-the-art codes find as well. The results coincide most 
commonly with those of the \texttt{AHF}, \texttt{Rockstar}, and \texttt{Subfind} halo finders in 
combination with the \texttt{MergerTree}, \texttt{TreeMaker}, \texttt{Sublink}, and 
\texttt{VELOCIraptor} tree builders. One notable difference in our result however is that we 
recover more extreme mass growths and losses as well as fluctuations for sub-haloes due to the way 
the sub-halo hierarchy is established by \phew, specifically in cases where a sub-halo and its 
subsub-halo switch their order in the hierarchy between snapshots.

%%%%%%%%%%%%%%%%%%%%%%%%%%%%%%%%%%%%%%%%%%%%%%%%%%

	% Don't change these lines
	\bsp	% typesetting comment
	\label{lastpage}
\end{document}